\documentclass[a4paper,11pt]{article}
\usepackage{graphicx}
\usepackage{amsmath}
\usepackage{amssymb}
\usepackage{overpic}
\usepackage{hyperref}

\usepackage{multicol,color}
\definecolor{darkred}{rgb}{0.7,0,0.5}
\definecolor{darkgreen}{rgb}{0,0.3,0}
\hypersetup{pdfborder={0 0 0},colorlinks=true,urlcolor=darkred,citecolor=blue,linkcolor=darkgreen}

\newcommand{\id}{\hbox{1\kern-.27em l}}

\newcommand{\sltwo}{\mathrm{SL}(2, \mathbb{R})}
\newcommand{\slthree}{\mathrm{SL}(3, \mathbb{R})}
\newcommand{\slfour}{\mathrm{SL}(4, \mathbb{R})}
\newcommand{\sotwoone}{\mathrm{SO}(2,1)}
\newcommand{\sothreeone}{\mathrm{SO}(3,1)}
\newcommand{\sotwotwo}{\mathrm{SO}(2,2)}

\makeatletter

\@addtoreset{equation}{section}
\makeatother

\begin{document}

\thispagestyle{empty}
\setcounter{page}{0}

{\hfill{\tt ULB-TH/09-34}}

\vspace{20mm}

\begin{center} {\bf \Large Some Algebraic Aspects of Half-BPS\\[5mm]
Bound States in M-Theory}

\vspace{18mm}

Laurent Houart, Axel Kleinschmidt and Josef Lindman H\"ornlund

\footnotesize
\vspace{.9 cm}

{\em Service de Physique Th\'eorique et Math\'ematique,\\
Universit\'e Libre de Bruxelles \& International Solvay Institutes\\ Campus Plaine C.P. 231, Boulevard du
Triomphe, B-1050 Bruxelles, 
Belgium}

\vspace{.4 cm}

 {\tt
\{lhouart,axel.kleinschmidt,jlindman\}@ulb.ac.be} 

\end{center}

\vspace {.7cm}

{\abstract

\noindent We revisit non-marginal half-BPS solutions of M-theory in the framework of the possible existence of an underlying E$_{11}$   Kac-Moody symmetry.
In this context, non-marginal BPS solutions of M-theory can be described as exact solutions  of the brane $\sigma$-model E$_{10}/K(\mathrm{E}_{10})$, extending results obtained earlier for marginal BPS solutions.  We uncover an elegant and simple algebraic structure underlying the bound states by looking at subalgebras embedded in  E$_{10}$. Furthermore, we show that the non-marginal BPS solutions can be obtained from the elementary marginal ones by the action of $K(\mathrm{E}_{10})$ transformations.}

\newpage
\tableofcontents
\setcounter{equation}{0}

\section{Introduction}

Supergravity solutions preserving some fraction of the supersymmetries of the theory are of central importance in understanding non-perturbative aspects of supergravity and string theory. They typically enjoy powerful non-renormalization properties that make them largely insensitive to the value of the coupling constant. At the same time, if they are viewed as space-time instantons from the string theory perspective, they are non-perturbative states that also play a crucial role in the string counting of microstates contributing to the black hole entropy~\cite{Strominger:1996sh}. Such solutions are termed BPS states and have been pivotal in understanding the web of string dualities  and the M-theory conjecture~\cite{Hull:1994ys,Witten:1995ex,Obers:1998fb,Argurio:1998cp}. The classification of all BPS states is incomplete but remarkable progress has been made over the last years, see for example~\cite{Gauntlett:2002fz,Gran:2007fu}. When one is looking for solutions with commuting Killing vectors one can make use of the powerful techniques of U-duality and solution generation via classical (or quantum) hidden symmetries to classify orbits of BPS states~\cite{Obers:1997kk,Bossard:2009at}.

In the present work, we address these issues from the point of view of the conjectured E$_{10}$ and E$_{11}$ Kac-Moody symmetries of M-theory~\cite{West:2001as,Damour:2002cu}. These symmetries are thought to be realized non-linearly when acting on the fields of maximal supergravity theories and BPS solutions in this framework have been analysed for example in~\cite{Englert:2003py,West:2004st,Cook:2004er,Englert:2007qb}. As shown in~\cite{Englert:2003py,Englert:2004ph}, one can construct a one-dimensional sigma model, called brane $\sigma$-model, with target being the coset E$_{10}/K(\mathrm{E}_{10})$ where $K(\mathrm{E}_{10})$ is a non-compact subgroup of E$_{10}$ containing the group $\mathrm{SO}(9,1)$.  This model is a consistent truncation of a one dimensional $\sigma$-model E$_{11}/K(\mathrm{E}_{11})$, where $K(\mathrm{E}_{11})$ is a non-compact subgroup of E$_{11}$ containing the group $\mathrm{SO}(10,1)$. In the truncation, the geodesic parameter is identified with a spacelike coordinate. Studying the null geodesics on this  E$_{10}/K(\mathrm{E}_{10})$ is supposed to provide information about  static solutions to $D=11$ supergravity depending on one spatial direction. This was exemplified in~\cite{Englert:2003py} by describing all the elementary half-BPS solutions of $D=11$ supergravity in this language. These are the M$2$ and M$5$ branes and the gravitational Kaluza-Klein momentum and monopole solutions. All these solutions must be smeared to one dimension to make the correspondence exact. It is noticeable that all these solutions correspond to the embedding of an $\mathrm{SL}(2,\mathbb{R})/\mathrm{SO}(1,1)$ subcoset into E$_{10}/K(\mathrm{E}_{10})$~\cite{Englert:2007qb}. Different embeddings give the various half-BPS solutions with different orientations. The null geodesic on $\mathrm{SL}(2,\mathbb{R})/\mathrm{SO}(1,1)$ is basically unique, reflecting the universal structure behind all these elementary half-BPS solutions.\footnote{A similar analysis can be performed for the cosmological version of the geodesic model~\cite{Damour:2002cu,Englert:2004ph,Kleinschmidt:2005gz}. In this case, the solutions depend on the time variable and embedding $\mathrm{SL}(2,\mathbb{R})/\mathrm{SO}(2)$ yields (non-supersymmetric) S-brane type solutions. More complicated subcoset configurations can be considered~\cite{Henneaux:2006gp}.} For the geodesic E$_{10}$ model there exists a `dictionary' ~\cite{Damour:2002cu,Damour:2004zy} that relates low lying generators of the infinite-dimensional algebra E$_{10}$ to fields of supergravity, allowing the direct map of null geodesics to solutions of supergravity. The E$_{10}$ model is formally integrable.

M-theory also admits more complicated half-BPS solutions than the elementary branes which can be considered as bound states of the elementary states~\cite{Izquierdo:1995ms,Costa:1996yb,Larsson:2001wt}.  Here, we investigate whether these more complicated half-BPS states also have a simple underlying algebraic structure. Our answer will be affirmative and has been partly anticipated by the related recent work~\cite{Cook:2009ri}. In \cite{Cook:2009ri} the E$_{11}$  framework of~\cite{West:2001as} is used in which one does not restrict to functions of a single spatial variable but formally associates coset elements depending on all space-time variables to supergravity configurations. One of the results of~\cite{Cook:2009ri} was that the so-callled dyonic membrane~\cite{Izquierdo:1995ms}, which is a bound state of an M$2$ brane delocalized inside an M$5$ brane, can be described by using three step operators of E$_{11}$.\footnote{Configurations involving multiple step operators have been considered also in~\cite{Brown:2004jb} in a slightly different context.} The elementary half-BPS solutions only require a single step operator. Our first observation is that the three step operators needed to describe the dyonic membrane actually form an $\slthree$ group inside E$_{10}$ (or E$_{11}$). As for the $\mathrm{SL}(2,\mathbb{R})$ subgroups discussed above for elementary half-BPS solutions there are, of course, many embeddings of $\slthree$ inside E$_{10}$ corresponding to different orientations and also to similar bound states between different elementary branes, e.g. an M$2$ inside a KK$6$ monopole. For the $\slthree/\sotwoone$ subcoset it turns out that null geodesics are again very constrained and the dictionary mentioned above gives directly all the bound states of this type. Similarly, one can consider the embedding of larger subgroups in E$_{10}$; the only other example that we treat in this paper is $\slfour$ which can be used to describe a composite state consisting of either three or four branes, depending on the choice of subgroup $\sothreeone$ or $\sotwotwo$ of $K(\mathrm{E}_{10})$. These also correspond to known solutions constructed previously using familiar solution generating techniques for eleven dimensional supergravity. For example all M-theory solutions we find can be derived from the bound state constructed in \cite{Larsson:2001wt} but our methods transcend those used in~\cite{Larsson:2001wt} in that they can also give bound states involving D$8$ branes. The reason is that there is not only a map from the null geodesic to solutions of $D=11$ supergravity \cite{Cremmer:1978km} but also to Romans massive IIA supergravity \cite{Romans:1985tz} as well as type IIB supergravity. The map to massive type IIA supergravity requires symmetries that go beyond the E$_8$ U-duality transformations. It is important to emphasize that once we have choosen a given embedding of a subalgebra $ \mathfrak{g}$ in $ \mathfrak{e}_{11}$, the bound states solution is uniquely determined by a particular harmonicity ansatz on the fields corresponding to the Cartan subalgebra of $ \mathfrak{g}$ and the null geodesic equations.

Another result reported in the present paper is that the bound state solutions can in fact be obtained from the elementary solutions by symmetry transformations with the group $K(\mathrm{E}_{10})$. For example, starting from an elementary M$2$ brane one can rotate with an element of $\sotwoone\subset K(\mathrm{E}_{10})$ to obtain the dyonic membrane of~\cite{Izquierdo:1995ms}. Similarly, one could perform rotations with $\sothreeone$ or $\sotwotwo$ to arrive at the more complicated bound states describable by $\slfour$ subgroups. This can be viewed as part of the general BPS-orbit techniques discussed for example in~\cite{Bossard:2009at} but here applied to subgroups of E$_{10}$. The simple reason that the $K(\mathrm{E}_{10})$ symmetry transformations map half-BPS solutions to half-BPS solutions, i.e. preserve the number of unbroken supersymmetries, lies in the fact that the Killing spinor equation is $K(\mathrm{E}_{10})$-covariant.

In a next step, one can consider intersecting branes~\cite{Tseytlin:1996bh,Argurio:1997gt}. These correspond to commuting subgroups of E$_{10}$~\cite{Englert:2004it}. For example, the intersection of three dyonic membranes considered in~\cite{Costa:1996yb} is described algebraically by a subgroup $\slthree \times \slthree \times \slthree \subset \mathrm{E}_{10}$. Similar and more complicated configurations can easily be deduced. These solutions preserve smaller fractions of supersymmetry.

Though we have not used our methods to generate genuine new half-BPS solutions, the structures we highlight here organize the plethora of half-BPS solutions and provide a simple algebraic framework for describing them. Spacetime brane configurations can be simply deduced from the algebraic structure of steps operators as we will explain in more detail below. Furthermore, complicated space-time bound state solutions can be derived by rotating simple geodesics using $K(\mathrm{E}_{10})$ symmetry transformations.

Our paper is organized as follows. In section~\ref{sec:e11}, we review the correspondence between the $\mathrm{E}_{10}$ geodesic model and supergravity and recall how the algebraic structure is used to encode branes and the appearance of $\mathrm{SL}(2,\mathbb{R})$ for elementary half-BPS solutions of $D=11$ supergravity. In section~\ref{sec:twobranes}, the embedding of $\slthree$ in E$_{10}$ is then considered and shown to lead to  half-BPS bound state solutions. The embedding of $\slfour$ is shown to correspond to more complicated bound state solutions, examples of which we display in section \ref{sec:slfour}. Appendices provide supplementary material on our conventions for the space-time supergravity theories and for $\mathfrak{sl}(n,\mathbb{R)}$ algebras.

\section{Branes in the E$_{11}$ $\sigma$-model}
\label{sec:e11}

\begin{figure}
\begin{center}
\begin{overpic}[scale=0.6]{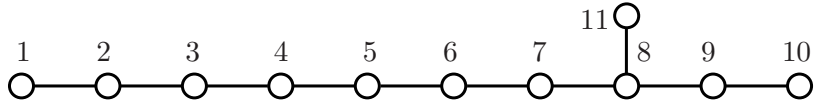}
\put(1,5){1}
\put(11,5){2}
\put(22,5){3}
\put(33,5){4}
\put(44,5){5}
\put(54,5){6}
\put(65,5){7}
\put(78,5){8}
\put(86,5){9}
\put(96,5){10}
\put(71,9){11}
\end{overpic}
\caption{\small The Dynkin diagram of $\mathfrak{e}_{11}$.}
\label{fig:E11dynkin}
\end{center}
\end{figure}

Let us now introduce the main tools to be used in this paper, the $\mathrm{E}_{11}$ $\sigma$-model and the supergravity/Kac-Moody dictionary. For this we will follow the description of the E$_{11}$ $\sigma$-model given in \cite{Englert:2003py}. We will also recall how to describe single extremal branes in this setting before considering their bound states in later sections. The $\mathfrak{e}_{11}$ algebra is a real Kac-Moody algebra defined by the Dynkin diagram in figure \ref{fig:E11dynkin} together with the usual Chevalley-Serre relations~\cite{Kac:book}. Counting how many times the simple 11th root shows up in a given root corresponding to a given step operator we get a $\mathbb{Z}$-grading
\begin{equation}
\label{eqn:grading}
\mathfrak{e}_{11} = \sum_{\ell\in \mathbb{Z}} \mathfrak{g}_\ell
\end{equation}
of $\mathfrak{e}_{11}$ such that  $\mathfrak{g}_0 = \mathfrak{gl}(11,\mathbb{R})$ and the positive and negative levels transform as $\mathfrak{gl}(11,\mathbb{R})$ representations. This corresponds to a so called level decomposition with respect to the regular $\mathfrak{gl}(11,\mathbb{R})$ subalgebra given by removing the $11$th node, in figure \ref{fig:E11dynkin} \cite{Damour:2002cu,Nicolai:2003fw, West:2002jj,Kleinschmidt:2003mf}. The lowest lying positive levels $\mathfrak{g}_{\ell}$ with $\ell=0,1,2,3$ are given in table \ref{tab:leve11}. 

\begin{table}[h!]
\begin{center}
\begin{tabular}{|c|c|c|}
\hline
$\ell$ &$ \mathfrak{sl}(11,\mathbb{R})$ Dynkin labels& Generator of $\mathfrak{e}_{11}$\\
\hline \hline
$0$ &$[ 1,0,0,0,0,0,0,0,0,1] $ & $K^a_{\ b}$\\
$1$ &$[ 0,0,0,0,0,0,0,1,0,0]  $ & $R^{\,a_1a_2a_3}$\\
$2$ &$[ 0,0,0,0,1,0,0,0,0,0]  $ & $R^{\, a_1...a_6}$\\
$3$ &$[ 0,0,1,0,0,0,0,0,0,1]  $ & ${R}^{\, a_0| a_1...a_8}$\\
\hline
\end{tabular}
\caption{\small \sl \small Level decomposition of $\mathfrak{e}_{11}$ under $\mathfrak{sl}(11,\mathbb{R})$ up to level $\ell=3$.}
\label{tab:leve11}
\end{center}
\end{table}

The operators at levels $\ell\geq1$ are $\mathfrak{gl}(11,\mathbb{R})$ tensors transforming according to the commutation relation
\begin{eqnarray}
\label{eqn:zerowithhigher}
[{K^a}_b, R^{c_1..c_n}] &=& {\delta^{c_1}}_b R^{ac_2...c_n}+...+{\delta^{c_i}}_b R^{c_1...a...c_n}+...+{\delta^{c_n}}_b R^{c_1...a} \nonumber \\
\end{eqnarray}
as the ${K^a}_b$ span the subalgbra $\mathfrak{g}_0$. The indices here are $\mathfrak{gl}(11,\mathbb{R})$ indices ranging from $1$ to $11$. From the grading (\ref{eqn:grading}) we have
\begin{equation}
[\mathfrak{g}_{\ell_1}, \mathfrak{g}_{\ell_2}] \subset \mathfrak{g}_{\ell_1+\ell_2}
\end{equation}
and in terms of the generators at level $1$,$2$ and $3$ this gives that
\begin{equation}
[R^{a_1a_2a_3},R^{a_4a_5a_6}] = R^{a_1...a_6}
\end{equation}
and
\begin{equation}
[R^{a_1a_2a_3},R^{a_4...a_9}]=R^{[a_1|a_2a_3]a_4...a_6} .
\end{equation}
Here $[a_1a_2a_3]$ implies anitsymmetrization with weight one. Now, to define the brane E$_{11}$ $\sigma$-model, we define an involution $\Omega$ of $\mathfrak{e}_{11}$, called the temporal involution~\cite{Englert:2003py}, such that the fixed point set of $\Omega$ is a subalgebra $\mathfrak{k}(\mathfrak{e}_{11})$. Its corresponding Lie subgroup of E$_{11}$, the group $K(\mathrm{E}_{11})$, is used to define the coset E$_{11}/K(\mathrm{E}_{11})$. We associate one of the $\mathfrak{gl}(11,\mathbb{R})$ indices with time.\footnote{We note that this does not always uniquely define the signature of space-time since the Weyl group does not commute with the involution and space-time signatures come in orbits of the Weyl group~\cite{Keurentjes:2004bv,deBuyl:2005it}. For example, the standard signature $(1,10)$ of $D=11$ supergravity is grouped together with the `exotic' signatures studied in~\cite{Hull:1998ym}. We always work in the standard Lorentzian theory in this paper.} The involution $\Omega$ fixing the $\mathfrak{k}(\mathfrak{e}_{11})$ subalgebra is now defined by
\begin{eqnarray}
\label{eqn:temporalinvolution1}
\Omega({K^a}_b) = -\epsilon_a \epsilon_b {K^b}_a
\end{eqnarray}
and
\begin{equation}
\label{eqn:temporalinvolution2}
\Omega(R^{a_1...a_n}) = -\epsilon_{a_1}...\epsilon_{a_n} R^{a_1...a_n}
\end{equation}
where $\epsilon_a = -1$ if $a$ is the chosen time index and $\epsilon_a  = +1$ otherwise. Using $\Omega$ we can form the projection operator 
\begin{equation}
\label{eqn:projection}
P_{\Omega} = \frac{1}{2}(\id +\Omega),
\end{equation}
projecting $\mathfrak{e}_{11}$ onto the  $\mathfrak{k}(\mathfrak{e}_{11})$ subalgebra. The complement $\mathfrak{p}(\mathfrak{e}_{11})$ is given by the image of $(\id-P_{\Omega})$ acting on $\mathfrak{e}_{11}$ so we get the vector space decomposition 
\begin{equation}
\label{eqn:e11decomposition}
\mathfrak{e}_{11} = \mathfrak{k}(\mathfrak{e}_{11}) \oplus \mathfrak{p}(\mathfrak{e}_{11}).
\end{equation}
At level 0 the generators 
\begin{equation}
k^{ab} = P_{\Omega} ({K^a}_b)
\end{equation}
in $\mathfrak{k}(\mathfrak{e}_{11}) $ generate a $\mathfrak{so}(10,1)$ algebra, so due to the fact that $\mathfrak{p}(\mathfrak{e}_{11})$ can be thought of as a $\mathfrak{k}(\mathfrak{e}_{11})$ representation, the tensors in $\mathfrak{p}(\mathfrak{e}_{11})$ are in particular tensors transforming under an eleven dimensional Lorentz group $\mathrm{SO}(10,1)$. We will see later that this Lorentz group can in fact be identified with the Lorentz group in space-time, when considering the so called dictionary in section \ref{sec:dictionary}.

Let $\mathcal{V}$ be a formal map from the real line, with coordinate $\xi$, to the infinite dimensional left coset E$_{11}/K(\mathrm{E}_{11})$. We write schematically this map as
\begin{equation}
\label{eqn:generalnu}
\mathcal{V} = \exp\left(\sum_{a,b} {\phi^b}_a {K^a}_b\right)\exp\left(\sum_ iC_i R^i\right)
\end{equation}
where $\sum_ iC_i R^i$ should be thought of as an infinite sum over the generators $R^i$ spanning the $\mathfrak{g}_{\ell}$ subspaces, with $\ell>0$ (given in table \ref{tab:leve11} up to level three), together with a matching $\xi$ dependent field $C_i = C_i(\xi)$. We will call the fields ${\phi^a}_a$ multiplying the diagonal in the ${K^a}_b$ matrix, {\it Cartan fields} as the ${K^a}_a$ elements generate a Cartan subalgebra of $\mathfrak{gl}(11,\mathbb{R})$. The fields $C_i$ we will denote as {\it coset-potentials}. The expression~(\ref{eqn:generalnu}) is gauge-fixed in a triangular (or Borel) gauge where only step operators of levels $\ell\geq 0$ have been used.\footnote{This gauge is not always admissible since the subgroup fixed by $\Omega$ is not the maximal compact subgroup~\cite{Englert:2007qb,Bossard:2009at}. For the particular BPS solutions we will consider this turns out not to be a problem.}

The $\sigma$-model Lagrangian is given by
\begin{equation}
\label{eqn:actione11}
\mathcal{L} = \eta^{-1}(\mathcal{P_{\xi}}| \mathcal{P_{\xi}}),
\end{equation}
where we have performed the splitting
\begin{equation}
\label{eqn:maurercartan}
\partial_{\xi} \mathcal{V}{\mathcal{V}}^{-1} = \mathcal{P}_{\xi} + \mathcal{Q}_{\xi}
\end{equation}
of the Maurer-Cartan form defined by $\mathcal{V}$ using the decomposition (\ref{eqn:e11decomposition}). The bilinear form $(.|.)$ is the canonical non-degenerate form on $\mathfrak{e}_{11}$ \cite{Kac:book}. The field $\mathcal{P}_{\xi}$ is the `velocity' vector, i.e.  the component of $\partial_{\xi} \mathcal{V}{\mathcal{V}}^{-1}$ along the coset and the `connection' $\mathcal{Q}_{\xi}$ is the component transverse to the coset. The index $\xi$ on $\mathcal{P}_{\xi}$ and $\mathcal{Q}_{\xi}$ indicate that these fields transform as vectors on the world line. The lapse function $\eta$ ensures reparameterization invariance of the Lagrangian (\ref{eqn:actione11}). There is locally always an affine parameterization such that $\eta=1$ and we will always choose this parameterization of the world line. The equations of motion derived from this action, under an infinitesimal change of the coset representative $\mathcal{V}$, is given by
\begin{equation}
\label{eqn:geodesiceqm}
 \partial_{\xi}  \mathcal{P}_{\xi}   - [\mathcal{Q}_{\xi} , \mathcal{P}_{\xi} ] = 0.
\end{equation}
From the variation of $\eta$ we get the quadratic Hamiltonian (or reparametrization) constraint
\begin{equation}
\label{eqn:null}
(\mathcal{P}_{\xi} | \mathcal{P}_{\xi} ) =0 .
\end{equation}
This system of equations exhibits two classes of symmetries, the first one being global E$_{11}$ invariance letting $\mathcal{V} \rightarrow \mathcal{V}g$. The other symmetry is a local $K(\mathrm{E}_{11})$ invariance, as $\mathcal{V}' = k(\xi) \mathcal{V}$, with $k:\mathbb{R} \rightarrow K(\mathrm{E}_{11})$, give an equivalent map to the coset. Under the local $K(\mathrm{E}_{11})$ the $\mathcal{P}_{\xi}$ vector transform as
\begin{equation}
\label{eqn:gaugetransfP}
\mathcal{P}_{\xi} \rightarrow k \mathcal{P}_{\xi} k^{-1} .
\end{equation}
It is with respect to this gauge-invariance that $\mathcal{Q}_{\xi} $ can be thought of as a connection since $\mathcal{Q}_{\xi}$ transform as
\begin{equation}
\label{eqn:gaugetransfQ}
\mathcal{Q}_{\xi} \rightarrow k \mathcal{Q}_{\xi} k^{-1} + \partial_{\xi}k k^{-1} .
\end{equation}
One can also identify $\mathcal{Q}_{\xi} $ with a Levi-Cevita connection on the coset. We note that in the fixed triangular gauge choice (\ref{eqn:generalnu}) one has to restore the gauge after a global right action by $g\in\mathrm{E}_{11}$ by means of a local and field-dependent compensating transformation in $\mathrm{K}(\mathrm{E}_{11})$. This induces a transformation of $\mathcal{P}_\xi$ and $\mathcal{Q}_\xi$ under the global $\mathrm{E}_{11}$.

The two equations (\ref{eqn:geodesiceqm}) and (\ref{eqn:null}) give that $\mathcal{V}$ describe a null-geodesic on the E$_{11}/K(\mathrm{E}_{11})$ coset. In terms of the map (\ref{eqn:generalnu}) the velocity vector $\mathcal{P}_{\xi} $ along the coset is expanded as 
\begin{eqnarray}
\mathcal{P}_{\xi}  &=& \frac12P_{\xi,a_1a_2} s^{a_1a_2} + \frac{1}{3!}P_{\xi,a_1a_2a_3} s^{a_1a_2a_3}+\frac{1}{6!}P_{\xi,a_1...a_6} s^{a_1...a_6} \nonumber \\
\label{eqn:cosetvelocity}
& &+\frac{1}{9!}P_{\xi,a_0|a_1...a_8}s^{a_0| a_1...a_8} + ...\, ,
\end{eqnarray}
where the $...$ indicate generators with level $\ell > 3$. Similarly the connection is given by
\begin{eqnarray}
\mathcal{Q}_{\xi}  &=& \frac12 Q_{\xi,a_1a_2} k^{a_1a_2} + \frac{1}{3!}P_{\xi,a_1a_2a_3} k^{a_1a_2a_3}+\frac{1}{6!}P_{\xi,a_1...a_6} k^{a_1...a_6}\nonumber \\
\label{eqn:connection}
& & +\frac{1}{9!}P_{\xi,a_0|a_1...a_8}k^{a_0| a_1...a_8} + ...\, .
\end{eqnarray}
Here, the $s^{a_1...a_n}$ generators are the projections of $R^{a_1...a_n}$ on the $\mathfrak{p}(\mathfrak{e}_{11})$ subspace, and the $k^{a_1...a_n}$ the projections onto $\mathfrak{k}(\mathfrak{e}_{11})$, i.e.
\begin{align}
s^{ab} &=  (1-P_{\Omega}) ({K^{a}}_{b}),&\quad
s^{a_1a_2a_3} &= (1-P_{\Omega}) (R^{a_1a_2a_3}) ,&\nonumber\\
s^{a_1...a_6} &= (1-P_{\Omega}) (R^{a_1...a_6}), &
s^{a_0| a_1...a_8} &= (1-P_{\Omega})( R^{a_0| a_1...a_8}),&
\end{align}
and
\begin{align}
k^{a_1a_2a_3} &= P_{\Omega} (R^{a_1a_2a_3}) ,&
k^{a_1...a_6} &= P_{\Omega} (R^{a_1...a_6}), &\nonumber\\
k^{a_0| a_1...a_8} &= P_{\Omega} (R^{a_0| a_1...a_8}).
\end{align}
We use the comma in the definitions of the $P$-fields in (\ref{eqn:cosetvelocity}) and (\ref{eqn:connection}) to separate different types of indices. Note that the $\xi$ -index does not transform under the subgroup $\mathrm{K}(\mathrm{E}_{11})$. In the dictionary, it will  be identified with a curved coordinate in space-time. The relations between the $P$-fields showing up in $\mathcal{P}_{\xi} $ and $\mathcal{Q}_{\xi} $ and the $C_i$ fields in (\ref{eqn:generalnu}) are given by generalized `covariant derivatives' \cite{Damour:2004zy}, thinking of the $P$-fields as the field strengths to the $C_i$ `potentials'. 

\subsection{The supergravity/Kac-Moody dictionary}
\label{sec:dictionary}

The E$_{11}$ $\sigma$-model described by the action (\ref{eqn:actione11}) leads to two different E$_{10}$ $\sigma$-model actions upon consistent truncation~\cite{Englert:2004ph}. 

The first one, named  cosmological E$_{10}$, is obtained by choosing the coordinate 1 to be the time coordinate and to put to zero all  the fields corresponding to step operators and the Cartan whose associated root contains $\alpha_1$'s , the simple root corresponding to the first node of figure \ref{fig:E11dynkin}. This gives a coset E$_{10}/\tilde K(\mathrm{E}_{10})$ where $\tilde K(\mathrm{E}_{10})$ contains $\mathrm{SO}(10)$ at level zero. In this model $\xi$ is identified to the time coordinate and permits to describe cosmological solutions in space-time.
The crucial information for describing the states in supergravity using the algebraic structure of $\mathfrak{e}_{10}$ is given by the so called supergravity/Kac-Moody dictionary. This dictionary tells us how to match the   E$_{10}$ fields in the coset, defined by (\ref{eqn:cosetvelocity}), with the space-time fields of eleven dimensional supergravity \cite{Damour:2004zy}. In the references \cite{Damour:2002cu,Damour:2004zy}  it is proven that the coset fields map directly to a truncated version of eleven dimensional supergravity. 

A second consistent truncation leading to the so-called brane E$_{10}$ $\sigma$-model   is obtained by first performing a Weyl reflection in the hyperplane perpendicular to $\alpha_1$ and then performing the same truncation. Because the non-commutativity of Weyl reflections and the temporal involution  $\Omega$ the timelike direction is now 2 and upon truncation one obtains another coset  E$_{10}/\  K(\mathrm{E}_{10})$ where $ K(\mathrm{E}_{10})$ contains $\mathrm{SO}(9,1)$ at level zero. Once the parameter $\xi$ is identified with a spacelike direction, this $\sigma$-model permits us to describe smeared static solutions of eleven dimensional supergravity \cite{Englert:2004ph}.

Thinking of the brane $\sigma$-model as a truncation of the E$_{11}$ $\sigma$-model we will surmise that a variant of the cosmological dictionary of~\cite{Damour:2002cu} also applies to the analysis presented in this paper. As we will only be concerned with half-BPS bound states, we will ignore the fermions and their corresponding dictionary. There is a significant difference, when working in the brane $\sigma$-model, compared to the discussion in \cite{Damour:2004zy}. In the brane $\sigma$-model we split the supergravity equations of motion with respect to a spatial coordinate and not with respect to a time-coordinate as in the Hamiltonian formalism. This splitting correspond to phrasing the eleven dimensional supergravity equations as a boundary-value problem instead of a initial conditions problem. 

The brane $\sigma$-model is based on the fact that, due to the temporal involution $\Omega$, the tensors in $\mathfrak{p}(\mathfrak{e}_{11})$ transform under a coset Lorentz group $\mathrm{SO}(10,1) \subset K(\mathrm{E}_{11})$. The dictionary now tells us that this Lorentz group can be identified with the Lorentz group in space-time.  Let us denote by $g_{\tilde{\mu} \tilde{\nu}}$ the metric in eleven dimensional supergravity and by ${e_{\tilde{\mu}}}^{\mu}$ its associated vielbein. Here tilded greek indices will be curved in space time, and plain greek indices will be flat. Although this is non-standard we have chosen these conventions to be able to separate $1$ and $\tilde{1}$, for example, where $1$ would be a specific flat coordinate, and $\tilde{1}$ the same but curved coordinate, so that ${e_{\tilde{1}}}^1$ would convert between them.\footnote{This is in a `zero shift' gauge, such that the metric is blockdiagonal, with $g_{\tilde{1}\tilde{1}}$ being one block.} The $P$-fields in (\ref{eqn:cosetvelocity}) can therefore be identified with fields in space-time, in an orthonormal local frame defined by the vielbein ${e_{\tilde{\mu}}}^{\mu}$. Hence we promote the coset $\mathfrak{gl}(11, \mathbb{R})$ indices $a,b,c$, on the tensors in $\mathfrak{p}(\mathfrak{e}_{11})$, to flat space-time indices, i.e.
\begin{equation}
P_{\xi, a_1...a_n} \rightarrow P_{\xi, \mu_1 ... \mu_n} .
\end{equation}
For smeared brane solutions the parameter $\xi$ is identified with the remaining transverse direction to the brane, and will be curved in space.

\subsubsection{The metric - level 0 and level 3}
\label{sec:metricdictionary}

First, let us discuss the dictionary between the space-time metric and the fields of $\mathfrak{e}_{11}$. As we will only be interested in a particular class of solutions, we will discuss a slightly simplified version of dictionary (compared to \cite{Damour:2004zy} for example), suitable for this class of solutions. This corresponds to setting all the fields ${\phi^a}_b$, defined in (\ref{eqn:generalnu}), with $a\neq b$ to zero, i.e. we only keep the Cartan fields $\phi^a \equiv {\phi^a}_a$ (and we define analogously $K_a \equiv {K^a}_a$). In space-time this corresponds to looking at solutions with no Kaluza-Klein momentum. Furthermore, the fields $\phi^a$ turn out to be associated with the diagonal components of the metric, according to the dictionary, in a way to be stated explicitly momentarily.
In the bound states we will consider, the only non-zero off-diagonal component of the space-time metric we will be concerned with is the spatial part of the so called Kaluza-Klein 1-form potential $A^{(\tilde{\mu}_n)}$,\footnote{The time component of the one-form of $A^{(\tilde{\mu}_n)}$ corresponds to the Kaluza-Klein wave, and the corresponding field shows up as a non-zero ${\phi^{a}}_b$ field in the dictionary, $a \neq b$. We will, however, not discuss bound states with KK-waves in this paper as mentioned above.} to be introduced below. The corresponding coset field will show up at a higher level in the dictionary. Let us therefore separate the diagonal and off-diagonal part of the space-time metric,
\begin{equation}
g_{\tilde{\mu} \tilde{\nu}} =  \hat{g}_{\tilde{\mu} \tilde{\nu}} + g^{\text{off}}_{\tilde{\mu} \tilde{\nu}},
\end{equation}
where $\hat{g}_{\tilde{\mu} \tilde{\nu}}$ is diagonal and $g^{\text{off}}_{\tilde {\mu} {\nu}}$ is zero on the diagonal. The dictionary now tells us that the diagonal components of the metric are given by
\begin{equation}
\label{eqn:branemetric}
\hat{g}_{\tilde{\mu} \tilde{\mu}} = e^{-2 \phi^a},
\end{equation}
where $\tilde\mu$ is the curved space-time index associated to $a$.
Let us now consider the off-diagonal parts of $g_{\tilde{\mu} \tilde{\nu}}$. Recall the Kaluza-Klein monopole metric (taking 9 to be time)
\begin{equation}
\label{eqn:kk6}
ds^2_{KK6} = H(\mathrm{d}x_{\tilde{1}}^2+\mathrm{d}x_{\tilde{2}}^2+\mathrm{d}x_{\tilde{3}}^2)+\mathrm{d}x_{\tilde{4}}^2+...-\mathrm{d}x_{\tilde{9}}^2+\mathrm{d}x_{\tilde{11}}^2+H^{-1}\big(\mathrm{d}x_{\tilde{10}}-A^{(\tilde{10})}\big)^2.
\end{equation}
The field $A^{(\tilde{10})}$ is the Kaluza-Klein potential we mentioned above and a 1-form on the three dimensional space $T'$ (spanned by $x_{\tilde{1}}$, $x_{\tilde{2}}$ and $x_{\tilde{3}}$) and $H$ is a harmonic function on $T'$. The coordinate $x_{\tilde{10}}$ is the curved coordinate on the euclidean Taub-NUT circle. The explicit form of $A^{(\tilde{10})}$ for the Kaluza-Klein monopole will be given explicitly when we discuss the KK6 in more detail below. For a general coordinate $x_{\tilde{\mu}_n}$ on the Taub-NUT circle we can from the `KK-potential' $A^{(\tilde{\mu}_n)}$ form the field strength $F_2^{(\tilde{\mu}_n)} = \mathrm{d}A^{(\tilde{\mu}_n)}$. Furthermore we can form the dual field strength 
\begin{equation}
\label{eqn:gravitationalduality}
F_9^{(\tilde{\mu}_n)} = \star F_2^{(\tilde{\mu}_n)}.
\end{equation}
Observe that $F_9^{(\tilde{\mu}_n)}$ is in general not a closed form, just as in the case of $F_7$. One may also remark that the dualization (\ref{eqn:gravitationalduality}) is with respect to the full metric $g_{\tilde\mu\tilde\nu}$ and may be non-linear, as the off-diagonal terms of the metric involve $A^{(\tilde{\mu}_n)}$. In the approach we will use below, this will however not be a problem as we will only be concerned with going the other way, from $F_9^{(\tilde{\mu}_n)}$ to $F_2^{(\tilde{\mu}_n)}$, always dualizing $F_9^{(\tilde{\mu}_n)}$ with the diagonal metric $\hat{g}_{\tilde{\mu} \tilde{\nu}}$ given by the correspondence (\ref{eqn:branemetric}). The dictionary now gives that $F_9^{(\tilde{\mu}_n)}$ is associated to a level $\ell=3$ component of $\mathcal{P}_\xi$. In terms of $F_2^{(\tilde{\mu}_n)}$, we have
\begin{equation}
\label{eqn:kkdictionary}
F^{(\nu_n)}_{\nu_1 \nu_2}  =\eta^{\nu_n \mu_n} \epsilon_{\nu_1 \nu_2\mu_1...\mu_9} {e_{\mu_1}}^{\xi}P_{\xi, {\mu}_n| \mu_2...\mu_9}.
\end{equation}
Observe that we must interpret the $(\nu_n)$ index on $F^{(\nu_n)}_{\nu_1 \nu_2} $ as flat, and hence when converting $F^{(\nu_n)}_{\nu_1 \nu_2} $ to curved indices, also this index needs to be converted. Furthermore, as the Taub-NUT circle necessary is one of the directions of $F^{(\nu_n)}_9$, the $\nu_n$ index will be the same as one of the $\mu_m$ indices, hence there is one repeated index on the mixed symmetry object $P_{\xi,\mu_n|\mu_1\ldots\mu_8}$. We will see an explicit example of how the dictionary for the Kaluza-Klein potential works when discussing the KK6 monopole in section \ref{sec:kk6}.

\subsubsection{The four-form - level 1 and level 2}

At level 1 and level 2 in the level decomposition of $\mathfrak{e}_{11}$ we find the four-form $F_4$ present in eleven dimensional supergravity and its dual $F_7$. If we identify $\xi$ with the curved $x_{\tilde{\nu}_1}$ coordinate, the components of $F_4$ with legs in $x_{\tilde{\nu}_1}$ direction are given by
\begin{equation}
\label{eqn:electricfourform}
F_{\nu_1 \mu_1\mu_2\mu_3} = {e_{\nu_1}}^{\xi}P_{\xi,\mu_1 \mu_2 \mu_3},
\end{equation}
with $P_{\xi,\mu_1 \mu_2 \mu_3}$ defined by (\ref{eqn:cosetvelocity}). Here, $F_{\nu_1 \mu_1\mu_2\mu_3} $ is given in flat indices. For the components of $F_4$ with no legs in the $\xi$ directions, we have the relation
\begin{equation}
\label{eqn:magneticfourform}
F_{\mu_1\mu_2\mu_3\mu_4} = \frac{1}{4!}\epsilon_{\mu_1\mu_2\mu_3\mu_4\nu_1...\nu_7}{e_{\nu_1}}^{\xi} P_{\xi, \nu_2 ...\nu_7} ,
\end{equation}
with $ P_{\xi, \nu_2 ...\nu_7} $ the coset field in $\mathfrak{p}(\mathfrak{e}_{11})$ at level two. Observe that in the case of a diagonal metric, we can integrate this correspondence to a matching of the potentials, so that
\begin{equation}
A_{\tilde{\mu_1}\tilde{\mu_2}\tilde{\mu_3}} = C_{a_1a_2a_3}
\end{equation}
and
\begin{equation}
A_{\tilde{\mu_1}...{\tilde{\mu_6}}} = C_{a_1...a_6}
\end{equation}
where now the tilded indices of the three-form potentials are curved in space-time. Here $C_{a_1a_2a_3}$ and $C_{a_1...a_6}$ are the coset potentials, defined by (\ref{eqn:generalnu}), and associated to the generators at level one and two. The supergravity forms $A_3$ and $A_6$ are the `electric' and `magnetic' potentials to $F_4$, defined by (\ref{eqn:electricpotential}) and (\ref{eqn:sevenform}). Observe that these two equations are not tensor equations and are not covariant under coordinate transformations in space-time.

\subsection{Embedding of subalgebras}
\label{sec:embedding}

As we will look at bound states of branes corresponding to a certain class of $\mathfrak{sl}(n, \mathbb{R})$ algebras, with $n >2$, we will here discuss how to embed such algebras in $\mathfrak{e}_{11}$. First, the Cartan elements of our subalgebras will always be embedded in the Cartan subalgebra of $\mathfrak{gl}(11,\mathbb{R})$, so that the associated fields will describe the diagonal parts of the eleven dimensional metric via (\ref{eqn:branemetric}). The positive step operators of the $\mathfrak{sl}(n, \mathbb{R})$ algebra will be embedded at higher levels and hence associated to for example components of the four-form or to the Kaluza-Klein potential. We will see that in all cases, $\mathfrak{sl}(n, \mathbb{R})$ generators embedded in $\mathfrak{e}_{11}$ as generators $R^{a_1...a_n}$ where one of the $a_i$ is a time-index, will be interpreted as a physical brane.

Regarding the $\mathrm{SL}(n,\mathbb{R})$ $\sigma$-models, the equations of motion (\ref{eqn:geodesiceqm}) and the quadratic constraint (\ref{eqn:null}) will apply to the coset representatives of maps into these subcosets of E$_{11}/K(\mathrm{E}_{11})$. This is analogous to the usual truncation of equations of motion to a subset of fields. If we always truncate to subalgebras, then this truncation is automatically consistent.

One can of course embed other subalgebras than $\mathfrak{sl}(n, \mathbb{R})$ in $\mathfrak{e}_{11}$. Furthermore one can embed the algebras at higher levels, as discussed in \cite{Englert:2007qb} in which a tower of subalgebras at higher levels are constructed by Weyl reflections of a $\mathrm{SL}(2,\mathbb{R})$ describing an extremal elementary brane. These would, in the E$_{11}$ approach to M-theory, corresponds to exotic states in a U-duality multiplet. None of these two questions will be addressed in this paper.

 \subsection{The $\sltwo/\mathrm{SO}(1,1)$ $\sigma$-model and half-BPS branes}
\label{sec:SL2cos}

To give the background before we consider how to describe the bound states in the $\mathrm{E}_{11}$ framework, we first recall the embedding of the (smeared versions of the) elementary M2 and M5 branes, and the KK6 monopole. A single extremal M-brane is characterized by a diagonal space-time metric and one non-zero component of $F_4$. The KK6 will be expressed in terms of the diagonal metric $\hat{g}_{\tilde{\mu} \tilde{\nu}}$ and a non-zero KK potential $A^{(\tilde{\mu}_n)}$. In $\mathrm{E}_{11}$ this corresponds to some non-zero Cartan fields at level zero and one non-zero field at a higher level, level one for M2, level two for M5, and level three for the KK6. The maps into the $\sltwo/\mathrm{SO}(1,1)$ coset, defining this single-brane $\sigma$-model are given by the coset representative \cite{Englert:2003py, West:2004st}
\begin{equation}\label{eqn:braneel}
\mathcal{V}_{\text{brane}} = \exp(\phi h)\exp (C e).
\end{equation}
Here, we take the $SL(2,\mathbb{R})$ generators to be $e,h,f$ with the standard commutation relations
\begin{equation}\label{eqn:SL2gens}
\left[h,e\right] = 2e\,,\quad \left[h,f\right]=-2f\,,\quad \left[e,f\right] = h\,.
\end{equation}
From (\ref{eqn:maurercartan}) and (\ref{eqn:cosetvelocity}) we find that 
\begin{equation}
P_{\xi} = \exp(2 \phi) \partial_{\xi} C,
\end{equation}
together with $\partial_{\xi} \phi$ are the two fields in the $\mathrm{SL}(2,\mathbb{R})$ coset velocity. The equations of motion, (\ref{eqn:geodesiceqm}), for $\mathcal{V}_{\text{brane}}$ to trace out a geodesic on the coset manifold $\sltwo/\mathrm{SO}(1,1)$ read
\begin{equation}
\partial_{\xi}^2 \phi + \frac{1}{2}\exp (4 \phi) (\partial_{\xi} C)^2 = 0
\end{equation}
and
\begin{equation}
\partial_{\xi}^2 C+4 \partial_{\xi} C \partial_{\xi} \phi = 0.
\end{equation}
Furthermore, we impose the quadratic constraint (\ref{eqn:null}),
\begin{equation}
(\partial_{\xi} \phi)^2-\frac{1}{4}\exp (4 \phi) (\partial_{\xi} C)^2 = 0
\end{equation}
so that the geodesic is light-like. The solution is then given by~\cite{Englert:2003py}
\begin{equation}
\label{eqn:harmonic}
\phi = \frac{1}{2} \log H
\end{equation}
and
\begin{equation}
\label{eqn:sltwosolution}
C = H^{-1},
\end{equation}
if $H = a+b \xi$, i.e. $H$ is a harmonic function in one variable, satisfying 
\begin{equation}\label{eqn:SL2harm}
\partial_\xi^2 H=0.
\end{equation}
For this solution, we find
\begin{equation}
\label{eqn:sl2P}
P_\xi = H \partial_{\xi}H^{-1} .
\end{equation}
We have put the integration constant in the $C(\xi)$  field to zero. When un-smearing the brane solutions all that one does is letting the functions $H$ be harmonic functions in the space transverse to the unsmeared brane solution. Let us now see how this simple $\sltwo$ model describe all the single extremal branes of M-theory, embedding the $\sigma$-model in different ways in E$_{11}$.

\subsubsection{The M2 brane}
\begin{table}[h!]
\begin{center}
\begin{tabular}{|c|cccccccc|c|cc|}
\hline
Branes &$1$& $2$ & $3$ & $4$ & $5$ & $6$ & $7$ & $8$ & $9$\ (t) & $10$ & $11$\\
\hline\hline
M2 &\, & \, & \, &\,& \, & \, &\, &\, &$\bullet$ &$\bullet$& $\bullet$ \\
\hline
\end{tabular}
\end{center}
\caption{\small \sl \small The space-time position of the single M2 brane. Here 9 is the time-direction, as indicated with $(t)$.}
\label{tab:m2}
\end{table}

First, consider a single membrane. Let $x_{\tilde{9}}$ be the time coordinate. If we take the M2 brane to be positioned in space-time as in table \ref{tab:m2} the space-time metric is given by
\begin{equation}
\label{eqn:m2metric}
\mathrm{d}s^2_{M2} =H^{-2/3}(-\mathrm{d}x_{\tilde{9}}^2+\mathrm{d}x_{\tilde{10}}^2+\mathrm{d}x_{\tilde{11}}^2)+H^{1/3}(\mathrm{d}x_{\tilde{1}}^2+... +\mathrm{d}x_{\tilde{8}}^2)
\end{equation}
and the four form field strength will be
\begin{equation}
\label{eqn:m2fieldstrength}
F_4 = \mathrm{d}H^{-1} \wedge \mathrm{d}x_{\tilde{9}}\wedge \mathrm{d}x_{\tilde{10}}\wedge \mathrm{d}x_{\tilde{11}} .
\end{equation}
Hence, if the creation operator $e$ of $\mathfrak{sl}(2, \mathbb{R})$ (see (\ref{eqn:braneel}) and (\ref{eqn:SL2gens})) is embedded as 
\begin{equation}
\label{eqn:sl2e}
e = R^{9 1011}
\end{equation}
and the Cartan element is embedded as 
\begin{equation}
\label{eqn:cartan}
h = -\frac{1}{3}(K_{1}+...+K_{8})+\frac{2}{3}(K_{9}+K_{10}+K_{11}) 
\end{equation}
we find that the dictionary tells us that
\begin{equation}
\label{eqn:m2dictionaryflat}
 F_{1 9 {10} 11} = {e_1}^{\xi}H \partial_{\xi} H^{-1} ,
\end{equation}
using (\ref{eqn:sl2P}), or in terms of curved indices
\begin{equation}
\label{eqn:m2dictionarycurved}
 F_{\tilde{1} \tilde{9} \tilde{10} \tilde{11}} = \partial_{\xi} H^{-1}.
\end{equation}
Here we have identified $\xi$ with $x_{\tilde{1}}$. Using the map (\ref{eqn:branemetric}) for the metric, together with the embedding of $h$ described by (\ref{eqn:cartan}) we reproduce (\ref{eqn:m2metric}). To reproduce the M2 four-form (\ref{eqn:m2fieldstrength}) we unsmear (\ref{eqn:m2dictionarycurved}) by letting $H$ be a harmonic function in the eight dimensional space transverse to the brane. This is possible by virtue of the harmonic equation (\ref{eqn:SL2harm}) satisfied by $H$. That the definitions (\ref{eqn:sl2e}) and (\ref{eqn:cartan}) of $h$ and $e$ give the right commutation relations, i.e. correspond to a $\mathfrak{sl}(2,\mathbb{R})$ subalgebra in $\mathfrak{e}_{11}$, is assured by relation (\ref{eqn:zerowithhigher}). 

\subsubsection{The M5 brane}

\begin{table}[h!]
\begin{center}
\begin{tabular}{|c|ccccc|ccc|c|cc|}
\hline
Branes &$1$& $2$ & $3$ & $4$ & $5$ & $6$ & $7$ & $8$ & $9\ (t)$ & $10$ & $11$\\
\hline\hline
M5 &\, & \, & \, &\, & \, &$\bullet$ &$\bullet$ &$\bullet$ &$\bullet$ &$\bullet$& $\bullet$ \\
\hline
\end{tabular}
\end{center}
\caption{\small \sl \small The space-time position of the single M5 brane. Here 9 is the time direction.}
\label{tab:m5}
\end{table}

In the same way we can describe the M5 solution, using another $\mathfrak{sl}(2, \mathbb{R})$. If the M5 would sit in space time as in table \ref{tab:m5}, the supergravity solution is given by
\begin{equation}
\label{eqn:m5metric}
\mathrm{d}s^2_{M5} =H^{2/3}(\mathrm{d}x_{\tilde{1}}^2+... +\mathrm{d}x_{\tilde{5}}^2)+H^{-1/3}(\mathrm{d}x_{\tilde{6}}^2+...+\mathrm{d}x_{\tilde{11}}^2)
\end{equation}
and
\begin{equation}
\label{eqn:m5fieldstrength}
F_4 = \star_T \mathrm{d}H
\end{equation}
where $T$ is the flat five dimensional transverse space to the M5. To reproduce this supergravity solution we embed the above given $\sltwo$ $\sigma$ model in $\mathfrak{e}_{11}$ by defining
\begin{equation}
e = R^{6 7 89 {10}11}
\end{equation}
together with a Cartan generator
\begin{equation}
h =  -\frac{2}{3}(K_{1}+...+K_{5})+\frac{1}{3}(K_{6}+...K_{11})  .
\end{equation}
Due to (\ref{eqn:branemetric}) this reproduces (\ref{eqn:m5metric}). Regarding the four-form, (\ref{eqn:magneticfourform}) tells us that
\begin{equation}
F_{167891011} =  {e_1}^{\xi} H \partial_{\xi} H^{-1},
\end{equation}
where we have used (\ref{eqn:sl2P}). After dualizing this seven-form to a four form using 
\begin{equation}
F_7 = \star F_4
\end{equation}
and converting it to curved indices we find
\begin{equation}
F_{\tilde{2}\tilde{3}\tilde{4}\tilde{5}} = \partial_{\xi} H.
\end{equation}
The unsmearing of this four-form gives exactly (\ref{eqn:m5fieldstrength}), as $\xi$ is again identified with the coordinate $x_{\tilde{1}}$.

\subsubsection{The KK6 monopole}
\label{sec:kk6}

\begin{table}[h!]
\begin{center}
\begin{tabular}{|c|ccc|ccccc|c|c|c|}
\hline
Branes &$1$& $2$ & $3$ & $4$ & $5$ & $6$ & $7$ & $8$ & $9$\, (t) & $10$\, (N) & $11$\\
\hline\hline
KK6 &\, & \, & \, &$\bullet$ & $\bullet$ &$\bullet$ &$\bullet$ &$\bullet$ &$\bullet$ &$\bullet$& $\bullet$ \\
\hline
\end{tabular}
\end{center}
\caption{\small \sl \small The space-time position of the single KK6 brane. Here 9 is the time direction and 10 is the Taub-NUT direction.}
\label{tab:kk6}
\end{table}

Apart from the M2 and the M5, M-theory contains also the Kaluza-Klein monopole, or for short the KK6. Let the KK6 be as in table \ref{tab:kk6}, with $10$ being the Taub-NUT direction. Then the eleven dimensional metric is given by equation (\ref{eqn:kk6}). The 1-form $A^{(\tilde{10})}$ is the KK potential on the transverse space $T'$, and is hence a form on $T'$. The associated field strength will have the form 
\begin{equation}
\label{eqn:KKtwoform}
F_2^{(\tilde{10})} = \star_{T'} \mathrm{d}H .
\end{equation}
As described in section \ref{sec:metricdictionary}, the field corresponding to the dual of $F^{(\tilde{\mu}_n)}_2$ show up at level three in the level decomposition of $\mathfrak{e}_{11}$. Hence, the corresponding $\mathrm{SL}(2,\mathbb{R})$ $\sigma$-model encoding the metric (\ref{eqn:kk6}) is given by the embedding
\begin{eqnarray}
h &=& -K_1-K_2-K_3+K_{10}, \\
e &=& R^{{10}|4...{10}11},
\end{eqnarray}
into $\mathfrak{e}_{11}$. The repeated index $10$ in the definition of $e$ indicates the Taub-NUT-direction. After again identifying $\xi$ with $x_{\tilde{1}}$, the dictionary now tells us that we have the relation 
\begin{equation}
F^{(10)}_{1456789{10}11} = {e_1}^{\xi}H\partial_{\xi} H^{-1}
\end{equation}
and this will be the only non-zero component of $F_9^{(10)}$. When dualizing this 9-form to  the 2-form $F^{(10)}_2$, we use the diagonal metric, described by the Cartan element $h$. Doing this we find 
\begin{equation}
F_{\tilde{2}\tilde{3}}^{(\tilde{10})} = \partial_{\xi} H
\end{equation}
and unsmearing this 2-form we find (\ref{eqn:KKtwoform}).

\section{Bound states of two branes}
\label{sec:twobranes}

The main observation of this paper, inspired by \cite{Cook:2009ri}, is the fact that, instead of just embedding $\mathfrak{sl}(2,\mathbb{R})$ algebras in $\mathfrak{e}_{11}$, we can embed bigger algebras, and in fact these will correspond to bound states of two or more branes. As different bound states are described by the same $\sigma$-model as in the case of single extremal branes, we will first look at a general $\slthree/\sotwoone$ $\sigma$-model. This model will for example describe a bound state between an M2 and an M5 brane.

\subsection{An $\slthree/\sotwoone$ $\sigma$-model}
\label{sec:slthree}

Recall that the $\mathfrak{sl}(3,\mathbb{R})$ algebra consists of two Cartan elements $h_1$ and $h_2$ and three positive step operators $e_1$, $e_2$ and $e_{12} = [e_1, e_2]$, and three negative step operators $f_1,f_2$ and $f_{12}=-[f_1,f_2]$. The restriction of the temporal involution of $\mathfrak{e}_{11}$ to the $\mathfrak{sl}(3,\mathbb{R})$ subalgebras we will be interested in is 
\begin{eqnarray}
\label{eqn:sl3involution}
\Omega(e_1) = f_1,& \Omega(e_2) = -f_2,
\end{eqnarray}
and hence the fixed subalgebra will be $\mathfrak{so}(2,1)$. Denote by $k_1,k_2$ and $k_{12} = [k_1,k_2]$ the three generators that span $\mathfrak{so}(2,1)$ and by $s_1, s_2,s_{12}$ the three generators that together with $h_1$ and $h_2$ span the complement $\mathfrak{p}$ in $ \mathfrak{sl}(3,\mathbb{R})$ (under the involution fixing $\mathfrak{so}(2,1)$). We hence have the vector space decomposition
\begin{equation}
\mathfrak{sl}(3,\mathbb{R}) = \mathfrak{so}(2,1) \oplus \mathfrak{p} .
\end{equation}
This decomposition of  $ \mathfrak{sl}(3,\mathbb{R})$ is described in more detail in appendix \ref{app:sl3r}. We can now describe the $\slthree/\sotwoone$ $\sigma$-model with the coset represenative
\begin{equation}
\label{eqn:coset}
\mathcal{V} = \exp\left(\phi_1 h_1 + \phi_2 h_2\right)\exp\left(C_1 e_1+C_2e_2 + C_3 e_{12}\right).
\end{equation}
The fields $\phi_1, \phi_2$ and the $C_i, i = 1,2,3$ depend on the parameter $\xi$ that parametrizes the one-dimensional transverse space of the smeared solution. From (\ref{eqn:coset}) we find that 
\begin{equation}
\label{eqn:Pexpression}
\mathcal{P}_{\xi} = \partial_{\xi} \phi_1 h_1+ \partial_{\xi} \phi_2 h_2+P_{\xi,1} s_1+P_{\xi,2} s_2+P_{\xi,3} s_{12}
\end{equation}
and
\begin{equation}
\label{eqn:Qexpression}
\mathcal{Q}_{\xi} = P_{\xi,1} k_1+P_{\xi,2} k_2+P_{\xi,3} k_{12}
\end{equation}
where
\begin{eqnarray}
\label{eqn:fieldB1}
P_{\xi,1} &=& \exp(2\phi_1-\phi_2) \partial_{\xi} C_1, \\
\label{eqn:fieldB2}
P_{\xi,2} &=& \exp(2\phi_2-\phi_1) \partial_{\xi} C_2,\\
\label{eqn:fieldB3}
P_{\xi,3} &=& \exp(\phi_1+\phi_2) \left(\partial_{\xi} C_3-\frac{1}{2}(C_2\partial_{\xi} C_1-C_1\partial_{\xi} C_2)\right).
\end{eqnarray}
Note that the exponential factors of $\phi_1$ and $\phi_2$ in front of the expressions for the $P_{\xi,i}$ can be thought of as the conversion factors between curved and flat coordinates. Furthermore, the subscripts $1,2$ or $3$ on the $P$-fields defined by (\ref{eqn:Pexpression}) are not coordinate indices but only labels. In terms of (\ref{eqn:Pexpression}) and (\ref{eqn:Qexpression}) we find that the equations of motion (\ref{eqn:geodesiceqm}) for $\phi_1,\phi_2$ and the $P_{\xi,i}$ become
\begin{eqnarray}
\label{eqn:phi1}
\partial_{\xi}^2 \phi_1+\frac{1}{2}P_{\xi,1}^2+\frac{1}{2}P_{\xi,3}^2 &=& 0, \\
\label{eqn:phi2}
\partial_{\xi}^2 \phi_2-\frac{1}{2}P_{\xi,2}^2+\frac{1}{2}P_{\xi,3}^2 &=& 0, \\
\label{eqn:B1}
\partial_{\xi} P_{\xi,1}-P_{\xi,2} P_{\xi,3} +P_{\xi,1}(2\partial_{\xi}\phi_1-\partial_{\xi}\phi_2) &=& 0, \\
\label{eqn:B2}
\partial_{\xi} P_{\xi,2}-P_{\xi,1} P_{\xi,3} +P_{\xi,2}(2\partial_{\xi}\phi_2-\partial_{\xi}\phi_1) &=& 0,
\end{eqnarray}
and
\begin{equation}
\label{eqn:B3}
\partial_{\xi} P_{\xi,3} + P_{\xi,3} (\partial_{\xi}\phi_1+\partial_{\xi}\phi_2) = 0.
\end{equation}
The quadratic constraint (\ref{eqn:null}) becomes
\begin{equation}
\label{eqn:quadratic}
(\partial_{\xi} \phi_1)^2+(\partial_{\xi} \phi_2)^2-\partial_{\xi} \phi_1 \partial_{\xi} \phi_2 -\frac{1}{4}P_{\xi,1}^2+ \frac{1}{4}P_{\xi,2}^2-\frac{1}{4}P_{\xi,3}^2 = 0 .
\end{equation}
This is a rather complicated system of non-linear equations and is difficult to integrate explicitly, but because of the constraining $\slthree$ symmetry it turns out to be integrable. However, to compare with bound states in space-time we will restrict to a subset of solutions. This subset contains solutions for which the metric is expressible in harmonic functions. Other solutions are not easily unsmeared and thus have no immediate interpretation. Hence they will not be considered in this paper. More explicitly, we know that $\phi_1$ and $\phi_2$ will appear in the metric, via (\ref{eqn:branemetric}). As ansatz we therefore set 
\begin{eqnarray}
\label{eqn:choiceofphi}
\phi_1 &=& \frac{1}{2} \log H, \\
\phi_2 &=& \frac{1}{2} \log \tilde{H}.
\end{eqnarray}
where the functions $H$ and $\tilde{H}$ are harmonic in one Cartesian coordinate, i.e. linear functions in $\xi$.\footnote{In the terminology of \cite{Bossard:2009at}, this ansatz is equivalent to looking at the smallest nilpotent orbit in $\mathfrak{sl}(3,\mathbb{R})$. There is one more nilpotent orbit, but as this orbit is ruled out by the harmonic ansatz we will not consider it here.} This is because, as we have discussed earlier, we describe bound states smeared down to one dimension in the $\mathfrak{e}_{11}$ $\sigma$-models. So, we let
\begin{equation}
\label{eqn:genericH}
H =  a+b \xi
\end{equation}
and
\begin{equation}
\label{eqn:generictildeH}
\tilde{H} = c+d  \xi .
\end{equation}
The choice of $a$ and $c$ are important for the value $\mathcal{V}$ takes at $\xi=0$. For a smeared solution this has no direct physical importance, but when unsmearing the brane solution corresponding to this $\sigma$-model, the value of $\mathcal{V}(0)$ will be related to the value of the fields at infinity, i.e. to asymptotic flatness. As (\ref{eqn:B3}) is integrable, one finds that
\begin{eqnarray}
P_{\xi,3} &=& A\exp(- \phi_1 -\phi_2)\\
&= & \frac{A}{\sqrt{H \tilde{H}}},
\end{eqnarray}
where $A$ is an integration constant. From this we can express $P_{\xi,1}^2$ and $P_{\xi,2}^2$ in terms of $H$ and $\tilde{H}$ and $A$, using (\ref{eqn:B1}) and (\ref{eqn:B2}). The constant $A$ is fixed by the quadratic constraint (\ref{eqn:quadratic}) to be
\begin{equation}
A = -\sqrt{\frac{d}{b}} \partial_{\xi} H
\end{equation}
and we end up with
\begin{eqnarray}
P_{\xi,1} &=& -\sqrt{\frac{\alpha}{b}} \frac{\ \partial_{\xi} H }{H \sqrt{\tilde{H}}},\\
P_{\xi,2} &=&  -\sqrt{\frac{\alpha}{d}}\frac{\ \partial_{\xi} \tilde{H}}{\tilde{H} \sqrt{H}}  , \\
\label{eqn:P3}
P_{\xi,3} &=&  -\sqrt{\frac{d}{b}} \frac{\partial_{\xi} H}{\sqrt{H\tilde{H} }}, 
\end{eqnarray}
where $\alpha = bc-ad$. Here we have fixed the signs in front of the expressions for the $P_i$ from (\ref{eqn:B1}) and (\ref{eqn:B2}). We finally find from (\ref{eqn:fieldB1}),(\ref{eqn:fieldB2}) and (\ref{eqn:fieldB3}) that
\begin{eqnarray}
\label{eqn:C1}
C_1 &=& \sqrt{\frac{\alpha}{b}}H^{-1},\\
\label{eqn:C2}
C_2 &=&\sqrt{\frac{\alpha}{d}} \tilde{H}^{-1},
\end{eqnarray}
and
\begin{equation}
\label{eqn:C3}
C_3 = \frac{1}{2 \sqrt{bd}}\big(\frac{b}{H}+\frac{d}{\tilde{H}}\big) .
\end{equation}

\subsection{The dyonic membrane}
\label{sec:dyonic}
\begin{table}[h!]
\begin{center}
\begin{tabular}{|c|ccccc|ccc|c|cc|}
\hline
Branes &$1$& $2$ & $3$ & $4$ & $5$ & $6$ & $7$ & $8$ & $9$\, (t) & $10$ & $11$\\
\hline\hline
M2 &\, & \, & \, &\,& \, & \, &\, &\, &$\bullet$ &$\bullet$& $\bullet$ \\
\hline
M5 &\, & \, & \, &\, & \, &$\bullet$ &$\bullet$ &$\bullet$ &$\bullet$ &$\bullet$& $\bullet$ \\
\hline
\end{tabular}
\end{center}
\caption{\small \sl \small The space-time positions of the two branes. The $x_{\tilde{\imath}}, \tilde{\imath}=\tilde{1},...,\tilde{5}$ coordinates parametrize the space $T$ transverse to the brane and 9 indicates time.}
\label{tab:braneconf1}
\end{table}

As a first example of how this $\sigma$-model reproduce certain bound states in M-theory we will consider the dyonic membrane. This solution to the supergravity equations of motion was found in \cite{Izquierdo:1995ms}, and is a bound state between an M2 and an M5 brane (see table \ref{tab:braneconf1}). The corresponding eleven dimensional space-time metric is given by
\begin{eqnarray}
\mathrm{d}s^2_{M2/M5} &= &H^{1/3}\tilde{H}^{1/3}(\mathrm{d}s_T^2) +H^{1/3}\tilde{H}^{-2/3}(\mathrm{d}x_{\tilde{6}}^2+\mathrm{d}x_{\tilde{7}}^2+\mathrm{d}x_{\tilde{8}}^2) \nonumber  \\
\label{eqn:spacetime}
& & +\ H^{-2/3}\tilde{H}^{1/3}(-\mathrm{d}x_{\tilde{9}}^2+\mathrm{d}x_{\tilde{10}}^2+\mathrm{d}x_{\tilde{11}}^2) .
\end{eqnarray}
Note that 9 is the time coordinate. We will denote the overall space transverse to the two branes by $T$, so that the metric on $T$ is just the flat metric $\mathrm{d}s_T^2= \mathrm{d}x_1^2+\mathrm{d}x_{2}^2+\mathrm{d}x_{3}^2+\mathrm{d}x_{4}^2+\mathrm{d}x_{5}^2$. The field strength is given by\footnote{Observe that our four-form differs from the one given in \cite{Izquierdo:1995ms}.}
\begin{eqnarray}
\label{eqn:fourform}
F_4& = &\cos(\beta)(\star_T \mathrm{d}H)+\sin(\beta)\mathrm{d}H^{-1}\wedge\mathrm{d}x_{\tilde{9}}\wedge\mathrm{d}x_{\tilde{10}}\wedge\mathrm{d}x_{\tilde{11}} \nonumber \\
& &-\tan(\beta)\mathrm{d}x_{\tilde{6}}\wedge\mathrm{d}x_{\tilde{7}}\wedge\mathrm{d}x_{\tilde{8}}\wedge\mathrm{d} \tilde{H}^{-1}.
\end{eqnarray}
Note that the $\star_T$ is the Hodge dual on the overall flat transverse space $T$. This solution is different from normal intersecting brane solutions as the Chern-Simons term is non-zero. Here, $H$ and $\tilde{H}$ are harmonic functions away from the singularity sitting at radius $r=0$ in the five-dimensional transverse space $T$, i.e. obeying 
\begin{eqnarray}
\label{eqn:harmonicity}
\mathrm{d} \star_T \mathrm{d}H &=& 0, \\
\label{eqn:harmonicity2}
\mathrm{d} \star_T \mathrm{d}\tilde{H} &=& 0,
\end{eqnarray}
for $r\neq 0$, and related by
\begin{equation}
\label{eqn:Hrelation}
 \tilde{H} = \sin^2(\beta)+H\cos^2(\beta).
\end{equation}
For this space-time solution both $H$ and $\tilde{H}$ go to $1$ at infinity, to assure asymptotic flatness of the solution. The harmonicity of $H$ ensures furthermore the Bianchi identity $\mathrm{d}F_4=0$. This solution is shown in table \ref{tab:braneconf1}, and is a two parameter solution, characterized by two charges $Q$, $\tilde{Q}$ and an interpolating angle $\beta$. We define these charges by the duals of the 1-forms $H$ and $\tilde{H}$, i.e.
\begin{equation}
\label{eqn:charge}
Q \equiv \int_{S_{\infty}^4} \star_T \mathrm{d}H\quad\quad\text{and}\quad\quad
\tilde{Q} \equiv \int_{S_{\infty}^4} \star_T \mathrm{d}\tilde{H},
\end{equation}
as by (\ref{eqn:harmonicity}) and (\ref{eqn:harmonicity2}) these charges are conserved. Relation (\ref{eqn:Hrelation}) give that the charges are related as $\tilde{Q} = \cos^2(\beta)Q$. There are two obvious limits of this dyonic solution, in one the M5-brane disappears, setting $\beta = \pi/2$, and in the other the M2 disappears, setting $\beta=0$. From (\ref{eqn:fourform}) we see that for the legs of the 4-form standing in the brane directions we have the corresponding potentials
\begin{eqnarray}
\label{eqn:threeforms}
A_{\tilde{9}\tilde{10}\tilde{11}} &=& \sin(\beta)H^{-1},\\
A_{\tilde{6}\tilde{7}\tilde{8}} &=& \tan(\beta)\tilde{H}^{-1}.
\end{eqnarray}
The dual $A_6$ form will also have one non-zero component living in the internal space, given explicitly by
\begin{equation}
\label{eqn:sixform}
A_{\tilde{6}\tilde{7}\tilde{8}\tilde{9}\tilde{10}\tilde{11}} = \frac{1}{2}\cos{\beta}\left(\tilde{H}^{-1}+\frac{1}{\cos^2(\beta)}H^{-1}\right).
\end{equation}
When deriving the expression for $A_6$ we have used the Chern-Simons term (see (\ref{eqn:sevenform})). These three explicit expression for parts of the potentials, encode the full form of the four-form, although they are of course in different `electromagnetic' frames. One also immediately see the resemblance with (\ref{eqn:C1}), (\ref{eqn:C2}) and (\ref{eqn:C3}). We will in fact find an exact matching below after we consider fixing $a,b,c$ and $d$. As the above solution describes an M2 brane dissolved in the bigger M5 brane one commonly uses the notation M2 $\subset$ M5, to concisely denote this solution. 

To reproduce this solution, from the $\sigma$-model solution given above, we choose the embedding of the $\mathfrak{sl}(3,\mathbb{R})$ described in section \ref{sec:slthree} in $\mathfrak{e}_{11}$ such that
\begin{eqnarray}
h_1 &=& -\frac{1}{3}(K_{1}+...+K_{8})+\frac{2}{3}(K_{9}+K_{10}+K_{11}) \nonumber, \\
h_2 &=& -\frac{1}{3}(K_{1}+...+K_{5}+K_{9}+K_{10}+K_{11})+\frac{2}{3}(K_{6}+K_{7}+K_{8}), \nonumber \\
\label{eqn:e11embedding}
e_1 &=& R^{9 {10}11},\\
e_2 &=& R^{6 7 8}\nonumber ,\\
e_{12} &=& [e_1,e_2] = -R^{6 7 89 {10}11} \nonumber.
\end{eqnarray}
If we choose 9 to be time, the temporal involution (\ref{eqn:temporalinvolution2}) reduce to the involution (\ref{eqn:sl3involution}). As we mentioned in section \ref{sec:embedding}, $e_1$ and $e_{12}$ will correspond to physical branes as their embeddings in $\mathfrak{e}_{11}$ are as generators with a time index. From the Kac-Moody/Supergravity dictionary we now find that $P_{\xi,1}$ and $P_{\xi,2}$ should be matched with components of the supergravity four-form $F_4$ and $P_{\xi,3}$ should be matched with a component of $F_7 = \star F_4$, i.e. using (\ref{eqn:electricfourform}) and (\ref{eqn:magneticfourform})  we find
\begin{eqnarray}
F_{19 10 11} &=& {e_1}^{\xi}P_{\xi,1},\nonumber\\
\label{eqn:dyonicformcomponents}
F_{1678} &=& {e_1}^{\xi}P_{\xi,2},\\
F_{167891011} &=& {e_1}^{\xi}P_{\xi,3},\nonumber
\end{eqnarray}
where all the components of the space-time forms are expressed in flat coordinates. Let us now lift this smeared solution, to the dyonic membrane. Comparing the general harmonic functions (\ref{eqn:genericH}) and (\ref{eqn:generictildeH}) with the ones for the dyonic membrane we see that to match we have to fix the four constants $a,b,c$ and $d$. Ensuring asymptotic flatness in the five dimensional transverse space to the dyonic membrane we put
\begin{eqnarray}
a &=& 1, \\
c &=& 1,
\end{eqnarray}
and to fulfill (\ref{eqn:Hrelation}) we put
\begin{eqnarray}
b &=& q, \\
d& =& \cos^2(\beta) q,
\end{eqnarray}
where $q$ is such that
\begin{equation}
\label{eqn:spacetimecharge}
Q = \mathrm{Vol}(S^4) q .
\end{equation}
This implies that $bc - ad = \sin^2(\beta)$. The lifted field components, given by (\ref{eqn:dyonicformcomponents}) then become
\begin{eqnarray}
F_{\mu_i9 10 11}  &=&{e_{\mu_i}}^{\tilde{\mu}_i} \sin(\beta) \frac{\ \partial_{\tilde{\mu}_i} H}{H \sqrt{\tilde{H}}},\\
F_{\mu_i678} &=&  {e_{\mu_i}}^{\tilde{\mu}_i}\tan(\beta)\frac{\ \partial_{\tilde{\mu}_i} \tilde{H}}{\tilde{H} \sqrt{H}}  , \\
\label{eqn:P3fixed}
F_{\mu_i67891011} &=&  -{e_{\mu_i}}^{\tilde{\mu}_i}\cos(\beta) \frac{\partial_{\tilde{\mu}_i} H}{\sqrt{H\tilde{H} }}, 
\end{eqnarray}
where we have promoted $\partial_{\xi}$ to $\partial_{\tilde{\mu}_i}$, as $H$ and $\tilde{H}$ are now harmonic functions in five dimensions and $x_{\tilde{\mu}_i}$ are the coordinates on this transverse space. Now, these three field components reproduce the four form (\ref{eqn:fourform}) for the dyonic membrane, dualizing (\ref{eqn:P3fixed})  (as this component is the dual of the first component in (\ref{eqn:fourform})) and converting the flat indices to curved indices. However, comparing the components of $A_3$ and $A_6$ given in (\ref{eqn:threeforms}) and (\ref{eqn:sixform}) with the $C_i$ potentials (\ref{eqn:C1}), (\ref{eqn:C2}) and (\ref{eqn:C3}), given the above choice of parameters $a,b,c$ and $d$, the matching is immediate.

We have thus shown that the dyonic membrane is exactly reproduced by a one-dimensional $\slthree/\sotwoone$ $\sigma$-model, embedded in $\mathrm{E}_{11}$. The dyonic membrane in E$_{11}$ was first discussed in \cite{Cook:2009ri}.

\subsection{The M2 with magnetic Kaluza-Klein charge}
\label{sec:M2KK6}

Our second example of how the above given $\slthree$ $\sigma$-model can be embedded in E$_{11}$ is given by an M2 brane with magnetic Kaluza-Klein charge. This configuration is T-dual to the dyonic membrane by double T-duality in ten dimensions. Let us perform this transformation to derive the form of the space-time fields for this supergravity solution. We will then describe the corresponding embedding in $\mathfrak{e}_{11}$. Consider the reduction of the dyonic membrane discussed in the previous section to ten dimensions, by reducing on the M-theory circle, here chosen to be the $x_{\tilde{10}}$. Recall that IIA contains the potentials $C_1$, $C_3$ together with their field strengths $G_2 = \mathrm{d}C_1$ and $G_4 = \mathrm{d}C_3$ and their dual potentials $C_5$ and $C_7$. After reduction we find a F1/D4 bound state in IIA with form fields
\begin{eqnarray}
H_3 &=& \sin (\beta) \mathrm{d}H^{-1}\wedge \mathrm{d}x_{\tilde{9}}\wedge \mathrm{d} x_{\tilde{11}}, \\
G_4 &=& \cos(\beta) \star_T \mathrm{d} H + \tan(\beta) \mathrm{d}\tilde{H}^{-1}  \wedge \mathrm{d}x_{\tilde{6}}\wedge \mathrm{d} x_{\tilde{7}}\wedge \mathrm{d}x_{\tilde{8}}
\end{eqnarray}
and with the dilaton
\begin{equation}
\phi = \log (H^{-1/2} \tilde{H}^{1/4}).
\end{equation}
Performing a double T-duality in the $4,5$ directions we find from the usual Buscher rules for type II supergravity \cite{Bergshoeff:1995as,Myers:1999ps}
that the new RR fields are
\begin{eqnarray}
\mathrm{d}{\tilde{C}_5} &=&  \tan(\beta) \mathrm{d}\tilde{H}^{-1}  \mathrm{d}x_{\tilde{4}}\wedge \mathrm{d}x_{\tilde{5}} \wedge \mathrm{d}x_{\tilde{6}}\wedge \mathrm{d} x_{\tilde{7}}\wedge \mathrm{d}x_{\tilde{8}}, \\
\mathrm{d}{\tilde{C}_1} &=& \cos(\beta) (\star_{T'} \mathrm{d} H),
\end{eqnarray}
where $T'$ is the flat three-dimensional space with coordinates $x_{\tilde{1}},x_{\tilde{2}}$ and $x_{\tilde{3}}$ and we use tilde to denote the T-duality transformed fields. The NS-NS field $H_3$ stays unchanged and the new dilaton is
\begin{equation}
\tilde{\phi} = \log (H^{-1/2} \tilde{H}^{-1/4}).
\end{equation}
Hence we have gone to a F1/D6 bound state in IIA which we now want to lift back to eleven dimensional supergravity. Its metric is given by
\begin{eqnarray}
\mathrm{d}s^2_{F1/D6} &=& \tilde{H}^{1/2}(\mathrm{d}x_{\tilde{1}}^2+\mathrm{d}x_{\tilde{2}}^2+\mathrm{d}x_{\tilde{3}}^2)+\tilde{H}^{-1/2}(\mathrm{d}x_{\tilde{4}}^2+...+\mathrm{d}x_{\tilde{8}}^2)\nonumber \\
& &+H^{-1}\tilde{H}^{1/2}(-\mathrm{d}x_{\tilde{9}}^2+\mathrm{d}x_{\tilde{11}}^2) .
\end{eqnarray}
Before we perform this lifting we have to dualize the $C_5$ form back to a $C_3$ form. One then finds, given that the lift back to eleven dimensions is
\begin{equation}
F_4 = G_4 + H_3 \wedge \mathrm{d}x_{\tilde{10}},
\end{equation}
that
\begin{eqnarray}
\label{eqn:m2kk6fieldstrength}
F_4 &=& \sin(\beta)\cos(\beta) H^{-1} (\star_{T'} \mathrm{d}H^{-1}) \wedge \mathrm{d}x_{\tilde{9}}\wedge \mathrm{d}x_{\tilde{11}}\nonumber\\
& & +\sin(\beta) \mathrm{d}H^{-1}\wedge \mathrm{d}x_{\tilde{9}} \wedge \mathrm{d}x_{\tilde{11}} \wedge (\mathrm{d}x_{\tilde{10}} - C_1) .
\end{eqnarray}
Lifting the metric we must take into account that the $C_1$ field now is non-zero as
\begin{equation}
\mathrm{ds}^2_{11} = e^{-2\tilde{\phi}/3}\mathrm{d}s_{10}^2+ e^{4\tilde{\phi}/3}(\mathrm{d}x_{\tilde{10}}-C_1)^2 .
\end{equation}
Hence the eleven dimensional metric becomes
\begin{eqnarray}
\label{eqn:m2kk6metric}
\mathrm{d}s^2_{M2/KK6} &=&  H^{1/3}\tilde{H}^{2/3}(\mathrm{d}x_{\tilde{1}}^2+\mathrm{d}x_{\tilde{2}}^2+\mathrm{d}x_{\tilde{3}}^2)+H^{1/3}\tilde{H}^{-1/3}(\mathrm{d}x_{\tilde{4}}^2+...+\mathrm{d}x_{\tilde{8}}^2)\nonumber \\
& &+H^{-2/3}\tilde{H}^{2/3}(-\mathrm{d}x_{\tilde{9}}^2+\mathrm{d}x_{\tilde{11}}^2)+H^{-2/3}\tilde{H}^{-1/3}(\mathrm{d}x_{\tilde{10}}-C_1)^2 .\nonumber \\
\end{eqnarray}
\begin{table}[h!]
\begin{center}
\begin{tabular}{|c|ccc|ccccc|c|c|c|}
\hline
Branes &$1$& $2$ & $3$ & $4$ & $5$ & $6$ & $7$ & $8$ & $9$\, (t) & $10$\, (N) & $11$\\
\hline\hline
M2 &\, & \, & \, &\,& \, & \, &\, &\, &$\bullet$ &$\bullet$& $\bullet$ \\
\hline
KK6 &\, & \, & \, &$\bullet$& $\bullet$ &$\bullet$ &$\bullet$ &$\bullet$ &$\bullet$ &$\bullet$& $\bullet$ \\
\hline
\end{tabular}
\end{center}
\caption{\small \sl \small The space-time positions of the two branes in the bound state between a M2 and a KK6. Here 9 is time and 10 is the Taub-NUT direction.}
\label{tab:braneconf2}
\end{table}

We have thus found the bound state of an M2 and a KK6, displayed in table \ref{tab:braneconf2}, or equivalently an M2 brane with magnetic Kaluza-Klein charge. Let us now see how we can describe it in $\mathfrak{e}_{11}$. As we have an off-diagonal piece in the metric, we know from the dictionary discussed in section \ref{sec:metricdictionary} that we have to include a generator at level 3. Hence we choose the embedding
\begin{eqnarray}
h_1 &=& -\frac{1}{3}(K_{1}+...+K_{8})+\frac{2}{3}(K_{9}+K_{10}+K_{11}), \nonumber \\
h_2 &=&  -\frac{2}{3}(K_{1}+K_{2}+K_{3}+K_{9}+K_{11})+\frac{1}{3}(K_{4}+...+K_{8}+K_{10}),  \nonumber \\
\label{eqn:e11embedding4}
e_1 &=& R^{9 {10}11},\\
e_2 &=& R^{4 5 6 7 8 10},\nonumber \\
e_{12} &=& [e_1,e_2] =  \frac{1}{3}R^{10|4 5 6 7 89 {10}11} \nonumber.
\end{eqnarray}
Again, through relation (\ref{eqn:branemetric}), we see that this choice of Cartan generators reproduce the diagonal part of the M2/KK6 metric, i.e. (\ref{eqn:m2kk6metric}) with $C_1 = 0$, choosing $\phi_1$ and $\phi_2$ as in (\ref{eqn:choiceofphi}). For the four-form the dictionary tells us that $P_{\xi,1}$ is again matched with the component $F_{1 9 10 11}$ as for the dyonic membrane. For the $P_{\xi,2}$ field, the dictionary gives
\begin{equation}
\label{eqn:magneticS5}
F_{14567810} = {e_1}^{\xi}P_{\xi,2}
\end{equation}
and to go the dual four form, we should dualize this component. Here, however the metric is no longer diagonal and the dualization is a little more complicated than in the dyonic case, so let us first look at the KK potential.
Here the dictionary (\ref{eqn:kkdictionary}) gives that
\begin{equation}
F_{23}^{(10)} = \frac{1}{\sqrt{H \tilde{H}}} {e_1}^{\tilde{1}} {e_{10}}^{\tilde{10}}\partial_{\tilde{1}} H \cos(\beta),
\end{equation}
using (\ref{eqn:P3}) for $P_{\xi,3}$. Let us now unsmear this $\sigma$-model solution. Transforming the non-zero component $F^{(\mu_n)}_2$ to curved indices we get
\begin{equation}
\label{eqn:kkfieldstrength}
F^{(\tilde{10})}_2 = \cos(\beta) \star_{T'} \mathrm{d} H
\end{equation}
i.e. we find the right KK potential for the full space-time metric, (\ref{eqn:m2kk6metric}). Now we can dualize (\ref{eqn:magneticS5}), yielding
\begin{equation}
F_{\mu_1\mu_2 9 10} =- \frac{\tilde{H}}{\sqrt{H}} \frac{\sin(2 \beta)}{2}\epsilon_{\mu_1\mu_2\nu_1} {e_{\mu_3}}^{\tilde{\nu}_1} \partial_{\tilde{\nu}_1} \tilde{H}^{-1} 
\end{equation}
in flat coordinates and 
\begin{eqnarray}
F_{\tilde{\mu}_1\tilde{\mu}_2\tilde{9}\tilde{11}} &=& \tan(\beta) H^{-1} \epsilon_{\tilde{\mu}_1\tilde{\mu}_2\tilde{\mu}_3} \partial_{\tilde{\nu}_1} H \nonumber \\
& & + 2 \sin(\beta) C_{[\tilde{\mu}_1} \partial_{\tilde{\mu}_2]} H^{-1},
\end{eqnarray}
in curved coordinates, where we have written $C_1 = C_{\tilde{\mu}_i} \mathrm{d}x_{\tilde{\mu}_i}$. Hence, these components, together with $F_{\tilde{\mu}_i \tilde{9}\tilde{10}\tilde{11}}$ give exactly (\ref{eqn:m2kk6fieldstrength}) and we see that the $\slthree/\sotwoone$ also exactly reproduce the M2 with magnetic Kaluza-Klein charge, given the embedding (\ref{eqn:e11embedding4}). From an algebraic perspective this is not surprising. The map from the dyonic $\slthree$ to this $\slthree$ can be done using a Weyl-transformation, and this Weyl transformation is equivalent to the U-duality map we performed in space-time to obtain the magnetically Kaluza-Klein charged membrane.

\subsection{The M5 with magnetic Kaluza-Klein charge}
\label{sec:M5KK6}

In the U-duality orbit of the dyonic membrane we also have an M5 brane with magnetic Kaluza-Klein charge, analogous to the previous case with the M2 brane. The algebraic embedding corresponding to this solution is exactly as for (\ref{eqn:e11embedding4}) now taking the $\mathfrak{sl}(3,\mathbb{R})$ generator $e_2$ to contain the time index, i.e. letting for example $x_{\tilde{4}}$ to be the time coordinate. The space-time configuration of the M5 and the KK6 is shown in table \ref{tab:braneconfM5KK6}. 
\begin{table}[h!]
\begin{center}
\begin{tabular}{|c|ccc|c|ccccc|c|c|}
\hline
Branes &$1$& $2$ & $3$ & $4\,(t)$ & $5$ & $6$ & $7$ & $8$ & $9$ & $10$\, (N) & $11$\\
\hline\hline
M5 &\, & \, & \, & $\bullet$  & $\bullet$   & $\bullet$ &$\bullet$  &$\bullet$  & \, &$\bullet$& \, \\
\hline
KK6 &\, & \, & \, &$\bullet$& $\bullet$ &$\bullet$ &$\bullet$ &$\bullet$ &$\bullet$ &$\bullet$& $\bullet$ \\
\hline
\end{tabular}
\end{center}
\caption{\small \sl \small The space-time positions of the two branes in the bound state between an M5 and a KK6. Here 4 is time and 10 is the Taub-NUT direction.}
\label{tab:braneconfM5KK6}
\end{table}

\subsection{A (D6,D8) bound state}
Another intersesting bound state in the orbit of the these M-theory bound states, is a bound state between a D6 and D8 brane, a classical solution of massive type IIA supergravity in ten dimensions. The dictionary of this theory is described in the papers \cite{Kleinschmidt:2004dy,Henneaux:2008nr}. For this section we will assume some familiarity with the paper \cite{Henneaux:2008nr}, although the reasoning is completely analogous to what we have done for eleven dimensional supergravity. At level four in $\mathfrak{e}_{11}$ we have a generator
\begin{equation}
R^{a_1|a_2|a_3...a_{12}} \in \mathfrak{g}_4,
\end{equation}
that will encode the information about the mass parameter in massive IIA, permitting us to describe the D8 brane \cite{West:2004st, Englert:2007qb} . Let us therefore choose the embedding of our by now familiar $\mathfrak{sl}(3, \mathbb{R})$ in $\mathfrak{e}_{11}$ as follows;

\begin{eqnarray}
h_1 &=& -\frac{1}{3}(K_{1}+K_4+...+K_9+K_{11})+\frac{2}{3}(K_{2}+K_{3}+K_{10}), \nonumber \\
\label{eqn:e11embedding51}
h_2 &=&  -K_1-K_2-K_3+K_{10} ,
\end{eqnarray}
for the two Cartan elements and 
\begin{eqnarray}
e_1 &=& R^{2 {3}10},\nonumber \\
\label{eqn:e11embedding52}
e_2 &=& R^{10|4567891011},\\
e_{12} &=& [e_1,e_2] =  \frac{1}{3}R^{10|10|2 3 4 5 6 7 89 {10}11} \nonumber,
\end{eqnarray}
for the three positive step operators. The corresponding metric, reduced to a string frame metric in ten dimensions give
\begin{eqnarray}
\mathrm{d}s_{10}^2 &=& \tilde{H}^{1/2}\big( \mathrm{d}x_{\tilde{1}}^2+H^{-1} (\mathrm{d}x_{\tilde{2}}^2+\mathrm{d}x_{\tilde{3}}^2) \\
& & + \tilde{H}^{-1}(\mathrm{d}x_{\tilde{4}}^2+...-\mathrm{d}x_{\tilde{9}}^2+\mathrm{d}x_{\tilde{11}}^2) \big) .
\end{eqnarray}
This reduction provide us with the dilaton
\begin{equation}
e^{3/4 \phi} = \tilde{H}^{-1} H^{-2/3} .
\end{equation}
The embedding (\ref{eqn:e11embedding51}) and (\ref{eqn:e11embedding52}) into $\mathfrak{e}_{11}$ provide us now with the following matching
\begin{eqnarray}
P_{\xi,2310} &= &P_{\xi,1}, \\
P_{\xi, 4567891011} &=& P_{\xi,2},
\end{eqnarray}
and 
\begin{equation}
P_{\xi, 234567891011} = P_{\xi,3},
\end{equation}
where $P_{\xi, i}$ as usual are the three fields in the $\sigma$-model of section \ref{sec:slthree}. Define
\begin{equation}
F_{10}= \star \, m 
\end{equation}
to be the dual of the mass parameter. Recall also that massive IIA contains, the field $H_3 = \mathrm{d}B_2$, a two form field strength $G_2$ and its dual $G_8 = \star G_2$. The dictionary of \cite{Henneaux:2008nr} now tells us that
\begin{eqnarray}
H_{1 2 3} &=&  {e_1}^{\xi}P_{\xi, 2 3 10} \\
G_{1 4 5 6 7 8 9 11}& =& {e_1}^{\xi} P_{\xi, 4 5 6 7 8 910 11}
\end{eqnarray}
and 
\begin{equation}
F_{123456789 11}  = {e_1}^{\xi} P_{\xi, 234567891011} .
\end{equation}
Here, we have again identified $\xi$ with $x_{\tilde{1}}$. The bound state of a D6 and a D8 is also  derived by massive T-duality in \cite{Singh:2001gt}. The bound state is summarized in table~\ref{tab:braneconfD6D8}.

\begin{table}[h!]
\begin{center}
\begin{tabular}{|c|cccccccc|c|c|}
\hline
Branes &$1$& $2$ & $3$ & $4$ & $5$ & $6$ & $7$ & $8$ & $9\,(t)$ & $11$ \\
\hline\hline
D6 &\, & \, & \, &$\bullet$& $\bullet$& $\bullet$ &$\bullet$  &$\bullet$  &$\bullet$ &$\bullet$\\
\hline
D8 &\, & $\bullet$ & $\bullet$ &$\bullet$& $\bullet$ &$\bullet$ &$\bullet$ &$\bullet$ &$\bullet$ &$\bullet$ \\
\hline
\end{tabular}
\end{center}
\caption{\small \sl \small The space-time positions of the two branes in the bound state between a D6 and a D8. Here 9 is time and we have reduced from 11 dimensions on $x_{\tilde{10}}$ as this is our M-theory circle.}
\label{tab:braneconfD6D8}
\end{table}

\subsection{The $\slthree$ $\sigma$-model solution as an orbit space}
\label{sec:slthreeorbit}

In the above sections we showed that the $\slthree/\sotwoone$ $\sigma$-model reproduces exactly some bound states smeared down to one dimension. Hence, we can use the global $\slthree$ invariance of the $\sigma$-model to act on these solutions. In this section we will therefore use this symmetry to investigate the $\slthree$-orbit space of the dyonic membrane. We can divide the group transformations into two cases. In the first case we look at elements $g \in \slthree$ outside of the $\sotwoone$ subgroup and in the second case we consider $g \in \sotwoone \subset \slthree$. Consider the value of $\mathcal{V}$ at $\xi=0$, 
\begin{equation}
p = \mathcal{V}(0) . 
\end{equation}
In the following discussion we will for simplicity always set $p = \id$. This corresponds to putting $a=1$ and $c=1$ in the expressions (\ref{eqn:genericH}) and (\ref{eqn:generictildeH}) for $H$ and $\tilde{H}$ and shifting the potentials $C_i$ to vanish at $\xi=0$. For the dyonic solution we made this particular choice of $a,c$ to conform with asymptotic flatness. This is because when lifting the smeared solution, $\xi = 0$ maps to the $r \rightarrow \infty$ limit. The elements outside of the $\mathrm{SO}(2,1)$ subgroup will now move around the point $p$. In terms of the space-time fields this corresponds to coordinate transformations and gauge transformations of the potential $A_3$ (or equivalently $A_6$) that disappear in the expression for the field strengths. Hence, with a general $\slthree$ transformation we can always choose $\mathcal{V}$ so that $\mathcal{V}(0) \rightarrow \id$.

 The second type of transformations concerns the elements in $\sotwoone$. These elements preserve the base point $p$, but act non-trivially on the constants $b,d$ and on the $P$-fields, namely the five integration constants of (\ref{eqn:geodesiceqm}). One can most easily see this by considering the fact that to preserve $\mathcal{V}(0)$ after acting with $k \in \mathrm{SO}(2,1)$ we perform a gauge transformation with $k^{-1}$. This gauge transfomation show up in the equations of motion, as these are not $\mathrm{SO}(2,1)$ invariant but transform covariantly, as $\mathcal{P}_{\xi}$ and $\mathcal{Q}_{\xi}$ transform by (\ref{eqn:gaugetransfP}) and (\ref{eqn:gaugetransfQ}) respectively. The $\sigma$-model quantity encoding the information of these integration constants, is most simply formulated using the conserved Noether current $\mathcal{J} = \mathcal{V} \mathcal{P} \mathcal{V}^{-1}$. If we write $H = 1+q \xi$, we have for the $\sigma$-model solution described in section \ref{sec:slthree}, that the Noether current at $\xi=0$ is
\begin{equation}
\mathcal{J}(q, \beta) = \left(\begin{array}{ccc}q & q \sin(\beta)  & -q \cos(\beta)   \\ -q \sin(\beta)   & -q \sin^2(\beta) & q \cos(\beta)\sin(\beta) \\ q \cos(\beta)   &  q \cos(\beta)\sin(\beta) & -q \cos^2(\beta)\end{array}\right) .
\end{equation}
Here we have used the standard matrix representation of $\mathfrak{sl}(3,\mathbb{R})$. The parameter $q$ is proportional to the space-time charge $Q$ given by (\ref{eqn:charge}), according to (\ref{eqn:spacetimecharge}). What we have to do to investigate the transformation properties of the solution is now to conjugate this matrix with elements in $\sotwoone$. This makes perfect sense as with $\mathcal{V}(0)=\id$, we have $\mathcal{J}(q,\beta) \in \mathfrak{p}$ and $\mathfrak{p}$ is a $\sotwoone$ representation under the adjoint action. Applying the compact one-parameter subgroup $k(\alpha) = \exp (-2\alpha k_2)$ of $\sotwoone$, we see that
\begin{equation}
k(\alpha)\mathcal{J}(q, \beta)k(\alpha)^{-1} = \mathcal{J}(q, \beta+\alpha).
\end{equation}
Consider now for example the dyonic membrane. Setting $\alpha = \pi/2-\beta$ for example, we get the pure M2 solution. It is hence possible to rotate the M2 into the full dyonic solution by applying an $\sotwoone$ transformation. The pure M2 correspond to a Noether current given by $\mathcal{J} = q(h_1+2s_{1})$. This $\mathfrak{p}$-element is in fact a highest weight vector, thinking of $\mathfrak{p}$ as a $\mathfrak{so}(2,1) \cong \mathfrak{sl}(2,\mathbb{R})$ representation (see again appendix \ref{app:sl3r} for details). Hence it is annihilated by one Lie algebra generator, and diagonal to a second generator whose corresponding 1-parameter subgroup rescale the charge parameter $q$. The $\sotwoone$ orbit space of the M2 solution, is hence the dyonic bound state we found in section \ref{sec:dyonic}, or equivalently the coset  $\sotwoone/\mathbb{R} \cong \mathbb{R} \times S^1$.

We also note that using the $SO(2,1)$ rotation $k(\alpha)$ we can actually identify the above solution of the $\slthree/\sotwoone$ $\sigma$-model as a solution of an $\mathrm{SL}(2,\mathbb{R})/SO(1,1)$ coset, placing it on the same footing as the elementary solutions discussed in section~\ref{sec:SL2cos}. Consider a rotated $\mathfrak{sl}(2,\mathbb{R})$ subalgebra embedded in $\mathrm{E}_{11}$ defined by
\begin{eqnarray}
\tilde{h} &=& h_1 + \sin^2(\alpha) h_2 + \cos(\alpha)\sin(\alpha) (e_2+f_2),\nonumber\\
\tilde{e} &=& \cos(\alpha) e_1 + \sin(\alpha) [e_1,e_2],\nonumber\\
\tilde{f} &=& \cos(\alpha) f_1  + \sin(\alpha) [f_1,f_2],
\end{eqnarray}
where all generators on the right hand side refer to those of the dyonic $\mathrm{SL}(3,\mathbb{R})$ in (\ref{eqn:e11embedding}). One can verify that this constitutes a skewly embedded $\mathrm{SL}(2,\mathbb{R})\subset\mathrm{E}_{11}$. Solving the associated coset model and using the dictionary one recovers exactly the dyonic membrane solution with $\alpha$ being shifted relative to the interpolating angle $\beta$. In this way, one can think of the dyonic membrane and all other half-BPS states considered in this paper as being associated to $\mathrm{SL}(2,\mathbb{R})$ subgroups of $\mathrm{E}_{11}$, possibly embedded skewly. We, however, prefer to present them in terms of regularly embedded subgroups $\mathrm{SL}(n,\mathbb{R})$ together with harmonic ans\"atze for the Cartan fields (and possibly other simplifying assumptions), allowing us to solve the associated geodesic problem on the coset. The advantage of using the regular embedding is that the dictionary is much simpler to apply and that this makes also the intersection rules more transparent. We will see this in the following section, in which we will consider precisely the intersection rules for bound states.

\subsection{Intersection rules}
\label{intersl3}

\begin{table}[h!]
\begin{center}
\begin{tabular}{|c|ccc|ccccc|c|cc|}
\hline
Branes &$1$& $2$ & $3$ & $4$ & $5$ & $6$ & $7$ & $8$ & $9$\, (t) & $10$ & $11$\\
\hline\hline
M2 $\subset$ M5 &\, & \, & \, &\,& \, &$\bullet$  &$\bullet$ &$\circ$ &$\circ$ &$\circ$& $\bullet$ \\
\hline
M2 $\subset$ M5 &\, & \, & \, &$\bullet$&$\circ$&\, &$\circ$ &\, &$\circ$ &$\bullet$& $\bullet$  \\
\hline
M2 $\subset$ M5 &\, & \, & \, &$\circ$&$\bullet$&$\circ$ &\, &$\bullet$ &$\circ$&\, & $\bullet$  \\
\hline
\end{tabular}
\end{center}
\caption{\small \sl \small The space-time positions of the three intersecting bound-states. White circles indicate where the M2 is localized and 9 is the time direction.}
\label{tab:intersecting}
\end{table}
In this section we will derive the intersection rules for solutions of the $\slthree$ $\sigma$-model solved in section \ref{sec:slthree}. From purely algebraic considerations using the structure of $\mathfrak{e}_{11}$ one can find conditions for possible intersections of bound states between two branes to be stable. Algebraically, the conditions for an intersecting brane solution to be a solution of the equations of motion, is expressed by the fact that the corresponding algebras must commute as subalgebras of $\mathfrak{e}_{11}$, as this implies that two sets of fields do not couple \cite{Englert:2004it}. Let us first consider the case of two intersecting dyonic membranes. For the level one generators to commute, they need at least one index in common, as their commutator is totally antisymmetric. However, from the form of the Cartans $h_i$, as given in (\ref{eqn:e11embedding}), if they share more than one index, the $h$ in one subalgebra will have a non-trivial commutator with the step operators of the other subalgebra. For example, take a second M2 brane lying in the $8\, 9\, {10}$ directions, then $h_1$ of (\ref{eqn:e11embedding}), corresponding to a first brane in the directions $9\,10\,11$, would have the following commutator with $R^{8 9 10}$
\begin{equation}
[h_1, R^{8 9 10}] = -R^{8 9 10}.
\end{equation}
This implies that the two dyonic membranes should intersect in exactly three spatial directions, four including time, in such a way that the M2's do not intersect at all. Hence we get the intersection rule
\begin{equation}
(\mathrm{M2} \subset \mathrm{M5}) \cap (\mathrm{M2} \subset \mathrm{M5}) = 3 .
\end{equation}
To conclude, taking the first M2 $\subset$ M5 to lie in the $6 7 8 9 {10}11$ direction, a suitable choice is to take the second dyonic membrane to lie in the $4 5 79{10}11$ direction. The first two rows in table \ref{tab:intersecting} illustrates this configuration. This will correspond to a second $\mathfrak{sl}(3, \mathbb{R})$ embedded in $\mathfrak{e}_{11}$ by 
\begin{eqnarray}
h'_1 &=& -\frac{1}{3}(K_{1}+K_2+K_3+K_4+K_6+K_8+K_9+K_{{10}})\nonumber \\
& & \quad\quad\quad\quad\quad+\frac{2}{3}(K_{5}+K_{7}+K_{11}), \nonumber \\
h'_2 &=& -\frac{1}{3}(K_{1}+K_2+K_3+K_5+K_6+K_7+K_{9}+K_{11})\nonumber \\
& & \quad\quad\quad\quad\quad+\frac{2}{3}(K_{4}+K_{8}+K_{{10}}), \nonumber \\
\label{eqn:e11embedding3}
e'_1 &=& R^{5 {7}9},\\
e'_2 &=& R^{4 10 {11}},\nonumber \\
e'_{12} &=& [e'_1,e'_2] = -R^{4 5 79 {10}11} \nonumber.
\end{eqnarray}
From these considerations it is just a matter of combinatorics to find a third dyonic membrane intersecting with the other two in exactly three spatial directions. In our algebraic language this is equivalent to saying that we can find three commuting $\mathfrak{sl}(3,\mathbb{R})$ algebras, whose corresponding $\sigma$-model would describe the intersection of three dyonic membranes. Hence the algebra $\mathfrak{g}$ given by
\begin{equation}
\mathfrak{g} = \mathfrak{sl}(3,\mathbb{R})\oplus\mathfrak{sl}(3,\mathbb{R})\oplus\mathfrak{sl}(3,\mathbb{R}),
\end{equation}
suitably embedded in $\mathfrak{e}_{11}$ correspond to a triple intersection of the corresponding branes which is 1/8 BPS in space-time. An example of such an intersection is given in table \ref{tab:intersecting}. When compared with the results of \cite{Costa:1996yb} we see that our algebraic considerations lead to completely equivalent results. One can furthermore deduce that we cannot add a fourth subalgebra still complying with the above rules.

\section{Bound states of three or more branes}
\label{sec:slfour}

There are several extensions of the $\slthree$ $\sigma$-model discussed above. What we will do in this section is to consider an $\slfour$ group having as subgroup the $\slthree$ described in section \ref{sec:slthree}. First there are two possible choices of non-compact subgroup to $\slfour$ whose subalgebra is fixed by the involution $\Omega$. We get either $\sothreeone$ or $\mathrm{SO}(2,2)$, depending on how we embed the $\slfour$ in E$_{11}$. The difference depends on whether the simple root, or equivalently which node in the Dynkin diagram of $A_3$ associated with time, is the middle node or one of the side nodes, as indicated in figures~\ref{fig:so31} and \ref{fig:so22}.

\subsection{An $\slfour/\mathrm{SO}(3,1)$ $\sigma$-model}

\label{sec:sl4sigmamodel}

\begin{figure}
\vspace{5mm}
\begin{center}
\begin{overpic}[scale=0.6]{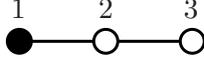}
\put(86,20){3}
\put(47,20){2}
\put(7,20){1}
\end{overpic}
\caption{\small Dynkin diagram for $A_3$, with the black node indicating time. In this case the fixed subalgebra under the temporal involution is $\mathfrak{so}(3,1)$. From the coloring of the nodes we get the involution (\ref{eqn:involutionso31}) by restriction of formula (\ref{eqn:temporalinvolution2}).}
\label{fig:so31}
\end{center}
\end{figure}

Let us first consider the case in which the temporal involution fixes a $\mathfrak{so}(3,1)$ subalgebra of $\mathfrak{sl}(4,\mathbb{R})$. The generators of $\mathfrak{sl}(4,\mathbb{R})$ are the three Cartan elements $h_1, h_2$ and $h_3$ and three simple step operators $e_1, e_2$ and $e_3$ and their commutators
\begin{eqnarray}
e_{12} = [e_1, e_2], & e_{23}=[e_2,e_3],  & e_{123} = [e_1 [e_2,e_3]]
\end{eqnarray}
together with six negative step operators. The temporal involution restricted to this class of $\mathfrak{sl}(4,\mathbb{R})$ subalgebras, and corresponding to the embedding defined by figure \ref{fig:so31}, is given by
\begin{eqnarray}
\label{eqn:involutionso31}
\Omega(e_1) = f_1,& \Omega(e_2)= -f_2,& \Omega(e_3) = -f_3,
\end{eqnarray}
as $e_1$ is associated with the node corresponding to the time coordinate. If we define $k_1,k_2,k_3,k_{12},k_{23}$ and $k_{123}$ to be the six generators spanning the $\mathfrak{so}(3,1)$ subalgebra of $\mathfrak{sl}(4,\mathbb{R})$, and let $s_1,s_2,s_3,s_{12}, s_{23}$ and $s_{123}$ together with $h_1,h_2$ and $h_3$ to span its complement $\mathfrak{p}'$ we have the reductive decomposition 
\begin{equation}
\label{eqn:reductivesl4so31}
\mathfrak{sl}(4,\mathbb{R}) = \mathfrak{so}(3,1) \oplus \mathfrak{p}' .
\end{equation}
For more details on $\mathfrak{sl}(4,\mathbb{R})$ and its reductive decomposition with respect to $\mathfrak{so}(3,1)$ see appendix \ref{app:sl4rso31}. Let us now look at the $\slfour/\sothreeone$  $\sigma$-model. The coset representative of maps into the $\slfour/\sothreeone$ coset manifold is given by
\begin{equation}
\mathcal{V} = \exp(\phi_1 h_1+\phi_2 h_2+\phi_3 h_3)\exp(C_1 e_1+C_2e_2+C_3 e_3+C_4 e_{12} +C_5 e_{23} +C_{6}e_{123}).
\end{equation}
Using the decomposition (\ref{eqn:reductivesl4so31}) and the usual splitting (\ref{eqn:maurercartan}) of the Maurer-Cartan form $\partial_{\xi} \mathcal{V} \mathcal{V}^{-1}$ we can write
\begin{eqnarray}
\label{eqn:Pexpression4}
\mathcal{P}_{\xi} &=& \partial_{\xi} \phi_1 h_1+ \partial_{\xi} \phi_2 h_2+ \partial_{\xi} \phi_3 h_3+P_{\xi,1} s_1+P_{\xi,2}  s_2+P_{\xi,3}s_3\nonumber \\
& &+P_{\xi,4}s_{12}+P_{\xi,5} s_{23}+P_{\xi,6} s_{123}
\end{eqnarray}
and
\begin{equation}
\label{eqn:Qexpression4}
\mathcal{Q}_\xi = P_{\xi,1} k_1+P_{\xi,2} k_2+P_{\xi,3} k_3+P_{\xi,4}k_{12}+P_{\xi,5} k_{23}+P_{\xi,6} k_{123}.
\end{equation}
The equations of motion of  the three Cartan fields for this model are
\begin{eqnarray}
\partial_{\xi}^2 \phi_1 + \frac{1}{2}\left(P_{\xi,1}^2+P_{\xi,4}^2+P_{\xi,6}^2\right)&=&0,\nonumber \\
\partial_{\xi}^2 \phi_2 + \frac{1}{2}\left(-P_{\xi,2}^2+P_{\xi,4}^2-P_{\xi,5}^2+P_{\xi,6}^2\right)&=&0, \\
\partial_{\xi}^2 \phi_3 + \frac{1}{2}\left(-P_{\xi,3}^2-P_{\xi,5}^2+P_{\xi,6}^2\right) &=&0\nonumber, 
\end{eqnarray}
and for the six coset field strengths
\begin{eqnarray}
\partial_{\xi} P_{\xi,1}-P_{\xi,2}P_{\xi,4}-P_{\xi,5}P_{\xi,6}+P_{\xi,1}(2\partial_{\xi}\phi_1-\partial_{\xi} \phi_2)&=&0, \nonumber\\
\partial_{\xi} P_{\xi,2}-P_{\xi,1}P_{\xi,4}-P_{\xi,3}P_{\xi,5}-P_{\xi,2}(\partial_{\xi}\phi_1-2\partial_{\xi}\phi_2+\partial_{\xi} \phi_3)&=&0,\nonumber\\
\label{eqn:slfoureqmsforPs1}
\partial_{\xi} P_{\xi,3}+P_{\xi,2}P_{\xi,5}-P_{\xi,4}P_{\xi,6}+P_{\xi,3}(2\partial_{\xi}\phi_3-\partial_{\xi} \phi_2) &=&0,\\
\partial_{\xi} P_{\xi,4}-P_{\xi,3}P_{\xi,6}+P_{\xi,4}(\partial_{\xi}\phi_1+\partial_{\xi} \phi_2-\partial_{\xi} \phi_3) &=&0,\nonumber\\
\partial_{\xi} P_{\xi,5}-P_{\xi,1}P_{\xi,6}-P_{\xi,5}(\partial_{\xi}\phi_1-\partial_{\xi} \phi_2-\partial_{\xi} \phi_3) &=&0,\nonumber\\
\partial_{\xi} P_{\xi,6}+P_{\xi,6}(\partial_{\xi}\phi_1+\partial_{\xi} \phi_3) &=&0.\nonumber
\end{eqnarray}
We also have the lapse constraint
\begin{eqnarray}
0 &= &(\partial_{\xi} \phi_1)^2-\partial_{\xi} \phi_1\partial_{\xi} \phi_2+(\partial_{\xi} \phi_2)^2-\partial_{\xi} \phi_2\partial_{\xi} \phi_3+(\partial_{\xi} \phi_3)^2 \nonumber \\
&&+\frac{1}{4}(-P_{\xi,1}^2+P_{\xi,2}^2+P_{\xi,3}^2-P_{\xi,4}^2+P_{\xi,5}^2-P_{\xi,6}^2) .
\end{eqnarray}
These equations cannot be solved directly from the ansatz we used in section \ref{sec:slthree} for the $\phi_i$ fields. Instead we will find a set of solutions with the approach of section \ref{sec:slthreeorbit}, i.e. by utilizing the $\slfour$ invariance of the $\sigma$-model to generate a bound state solution from a single brane.

\subsubsection{The dyonic membrane in a KK6}
\label{sec:dyonickk6}
\begin{table}[h!]
\begin{center}
\begin{tabular}{|c|ccc|ccccc|c|c|c|}
\hline
Branes &$1$& $2$ & $3$ & $4$ & $5$ & $6$ & $7$ & $8$ & $9$\, (t) & $10$\, (N) & $11$\\
\hline\hline
M2 &\, & \, & \, &\,& \, & \, &\, &\, &$\bullet$ &$\bullet$& $\bullet$ \\
\hline
M5 &\, & \, & \, &\,& \, & $\bullet$ &$\bullet$  &$\bullet$ &$\bullet$ &$\bullet$& $\bullet$ \\
\hline
KK6 &\, & \, & \, &$\bullet$& $\bullet$ &$\bullet$ &$\bullet$ &$\bullet$ &$\bullet$ &$\bullet$& $\bullet$ \\
\hline
\end{tabular}
\end{center}
\caption{\small \sl \small The space-time positions of the three branes in the bound state between an M2, an M5 and a KK6. Here 9 is time and 10 is the Taub-NUT direction.}
\label{tab:braneconf3}
\end{table}

\begin{figure}[!]
\begin{center}
\begin{overpic}[scale=0.8]{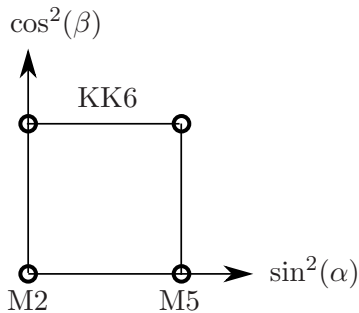}
\put(90,4){$\sin^2(\alpha)$}
\put(1,90){$\cos^2(\beta)$}
\put(0,-6){M2}
\put(52,-6){M5}
\put(24,64){KK6}
\end{overpic}
\caption{\small \small Different limits of the (M2,M5,KK6) bound state. For $\cos^2(\beta)=1$ the $\alpha$ dependence of the solution disappears and we reduce to the pure KK6 for all $\alpha$. For $\cos^2(\beta) = 0$ we interpolate between a M2 and a M5 (the solution in section \ref{sec:dyonic}), and when $\sin^2(\alpha) = 1$ we interpolate between an M5 and a KK6 (the solution in section \ref{sec:M5KK6}), and between an M2 and a KK6 when $\sin^2(\alpha)=0$). The area within the square is the full bound state, with varying amounts of M2, M5 and KK6 charge.}
\label{fig:braneinterpolation1}
\end{center}
\end{figure}

One way to embed an $\slfour$ in E$_{11}$ is to embed the `dyonic' $\mathfrak{sl}(3, \mathbb{R})$ in an $\mathfrak{sl}(4, \mathbb{R})$ algebra, such that the highest $\mathfrak{sl}(4, \mathbb{R})$ root shows up at level three in $\mathfrak{e}_{11}$. This will correspond to generating a dyonic membrane with magnetic Kaluza-Klein charge (see table \ref{tab:braneconf3}). We hence extend the embedding (\ref{eqn:e11embedding}) with the following $\mathfrak{sl}(2, \mathbb{R})$-subalgebra,
\begin{eqnarray}
h_3 &=& -\frac{1}{3}(K_{1}+K_2+K_3+K_6+K_7+K_8+K_9+K_{11})\nonumber \\
& &\quad\quad\quad\quad\quad  +\frac{2}{3}(K_{4}+K_{5}+K_{10}), \nonumber \\
\label{eqn:e11embedding2}
e_3 & = &R^{4 5 {10}} ,\\
f_3 &=&R_{4 5 {10}},\nonumber
\end{eqnarray}
generating together with our previous generators (also given in (\ref{eqn:e11embedding}))
\begin{eqnarray}
h_1 &=& -\frac{1}{3}(K_{1}+...+K_{8})+\frac{2}{3}(K_{9}+K_{10}+K_{11}) \nonumber, \\
h_2 &=& -\frac{1}{3}(K_{1}+...+K_{5}+K_{9}+K_{10}+K_{11})+\frac{2}{3}(K_{6}+K_{7}+K_{8}), \nonumber \\
\label{eqn:e11repeat}
e_1 &=& R^{9 {10}11},\\
e_2 &=& R^{6 7 8}\nonumber ,\\
e_{12} &=& [e_1,e_2] = -R^{6 7 89 {10}11} \nonumber,
\end{eqnarray}
an $\mathfrak{sl}(4, \mathbb{R})$, embedded in $\mathfrak{e}_{11}$, such that $e_1,e_2,e_3$ and $e_{12}$ together with 
\begin{eqnarray}
e_{23} & = & [e_2,e_3] = R^{4 5 6 7 8  {10}}, \nonumber \\
e_{123} & = & [e_1,[e_2,e_3]] = \frac{1}{3}R^{{10} | 4 ...{10}11},\nonumber
\end{eqnarray}
become the six creation operators of the algebra. We can generate this KK charged dyonic membrane, by starting with a single brane configuration and apply a $\sothreeone$ group transformation. Precisely as in the case of the dyonic membrane, the single brane configuration will correspond to a highest weight in $\mathfrak{p}'$ considered as a $\mathfrak{so}(3,1)$ representation, and hence only the three compact generators of $\mathfrak{so}(3,1)$ will act non-trivially on the brane. Again, for details on the description of $\mathfrak{p}'$ as a $\mathfrak{so}(3,1)$ representation, see appendix \ref{app:sl4rso31}. Hence we act with a general $\mathrm{SO}(3)$ element on our brane to generate the KK charged dyonic membrane. The $\mathfrak{so}(3)$ subalgebra is spanned by the three generators $k_2, k_3$ and $k_{23}$, and one of these generators leave our brane invariant. This will imply that our generated solution will depend on two angles, and not on three. Starting with the KK6 brane, corresponding to the vector $\lambda = h_1+h_2+h_3+2s_{123} \in \mathfrak{p}'$, we see that $\lambda$ commutes with $k_2$ and we end up with a two-angle solution with angles $\beta$ and $\alpha$, generated by $k_3$ and $k_{23}$. More concretely, acting with a general  $\mathrm{SO}(3)$ element on the KK6 we find the following $\sigma$-model solution. Defining 
\begin{equation}
\phi_i = \frac{1}{2} \log H_i
\end{equation}
we find that
\begin{eqnarray}
P_{\xi,91011} &=& \sin(\beta) \cos(\alpha) \frac{\partial_{\xi} H_1}{H_1 \sqrt{H_2}}, \nonumber\\
\label{eqn:PsInSL41}
P_{\xi,678} &=& \tan(\alpha) \frac{\partial_{\xi}H_2}{H_2 \sqrt{H_1 H_3}},\\
P_{\xi,4510}& =& \tan(\beta) \sin(\alpha) \frac{\partial_{\xi} H_3}{H_3 \sqrt{H_2}},\nonumber
\end{eqnarray}
and
\begin{eqnarray}
P_{\xi,67891011} & = &\sin(\beta)\sin(\alpha) \frac{\partial_{\xi} H_1}{\sqrt{H_1 H_2 H_3}}, \nonumber \\
\label{eqn:PsInSL42}
P_{\xi,4567810} & = &\tan(\beta)\cos(\alpha) \frac{\partial_{\xi} H_3}{\sqrt{H_1 H_2 H_3}},\\
P_{\xi,10|4567891011} & = &\cos(\beta) \frac{\partial_{\xi} H_1}{\sqrt{H_1 H_3}} \nonumber.
\end{eqnarray}
Here we have the relations
\begin{eqnarray}
H_1 &=& H, \\
H_2 &=& H_3 + \sin^2(\alpha) (H_1-H_3) , \\
H_3 &=& \sin^2(\beta)+\cos^2(\beta)H.
\end{eqnarray}
We immediately see that the level three field $P_{\xi,6}$ is identical to the KK charged M2 described in section \ref{sec:M2KK6} and hence the KK-potential is the same, i.e. the Kaluza Klein potential obeys (\ref{eqn:kkfieldstrength}). The full space time metric is therefore 
\begin{eqnarray}
\mathrm{d}s^2_{M2/M5/KK6} &=& H_1^{1/3}H_2^{1/3}H_3^{1/3}\big( \mathrm{d}x_{\tilde{1}}^2+\mathrm{d}x_{\tilde{2}}^2+\mathrm{d}x_{\tilde{3}}^2 \nonumber \\
&& +H_3^{-1}(\mathrm{d}x_{\tilde{4}}^2+\mathrm{d}x_{\tilde{5}}^2) \nonumber \\
&&+H_2^{-1}(\mathrm{d}x_{\tilde{6}}^2+\mathrm{d}x_{\tilde{7}}^2+\mathrm{d}x_{\tilde{8}}^2)\\
&& +H_1^{-1}(-\mathrm{d}x_{\tilde{9}}^2+\mathrm{d}x_{\tilde{11}}^2)\nonumber \\
&& +H_1^{-1}H_3^{-1}(\mathrm{d}x_{\tilde{10}}^2-A^{(\tilde{10})}) \big)\nonumber .
\end{eqnarray}
Here we have used the definition of the Cartan elements in (\ref{eqn:e11embedding2}) and (\ref{eqn:e11repeat}), and the metric dictionaries (\ref{eqn:branemetric}) and (\ref{eqn:kkdictionary}). It is now straightforward to derive the four-form from the six $P$-fields in (\ref{eqn:PsInSL41}) and (\ref{eqn:PsInSL42}) and we find
\begin{eqnarray}
F_4 &=& \sin(\beta) \cos(\alpha) \mathrm{d}H_1^{-1}\wedge \mathrm{d}x_{\tilde{9}}\wedge \mathrm{d}x_{\tilde{11}}\wedge(\mathrm{d}x_{\tilde{10}}-A^{(\tilde{10})}) \nonumber \\
& &-\tan(\beta) \mathrm{d}H_2^{-1}\wedge \mathrm{d}x_{\tilde{6}} \wedge \mathrm{d}x_{\tilde{7}} \wedge \mathrm{d}x_{\tilde{8}} \nonumber \\
& &+ \sin(\alpha) \tan(\beta) \mathrm{d}H_3^{-1}\wedge \mathrm{d}x_{\tilde{4}}\wedge \mathrm{d}x_{\tilde{5}}\wedge(\mathrm{d}x_{\tilde{10}}-A^{(\tilde{10})})\\
& &+\cos(\alpha) \tan(\beta) H_1^{-1} (\star_{T'} H_3)\wedge \mathrm{d}x_{\tilde{9}} \wedge \mathrm{d}x_{\tilde{11}} \nonumber \\
& &+\sin(\alpha) \sin(\beta) H_3^{-1} (\star_{T'} H_1)\wedge \mathrm{d}x_{\tilde{4}} \wedge \mathrm{d}x_{\tilde{5}} \nonumber,
\end{eqnarray}
where $\star_{T'}$ is the Hodge star on the flat three dimensional space with coordinates $x_{\tilde{1}},x_{\tilde{2}}$ and  $x_{\tilde{3}}$. One immediately see that $F_4$ is closed. This solution correspond to a bound state between an M2, an M5 and a Kaluza-Klein monopole, or equivalently a dyonic membrane with a magnetic Kaluza-Klein charge. By varying the angles $\alpha$ and $\beta$ we retrieve the pure brane limits as well as three types of bound states, namely the ones discussed in section \ref{sec:dyonic}, \ref{sec:M2KK6} and \ref{sec:M5KK6}. These limits are indicated in figure \ref{fig:braneinterpolation1}.

\subsubsection{An (F1,D6,D8) bound state}
\label{sec:f1d6d8}

We now turn to the description of half-BPS states of massive Type IIA in the coset $\slfour/\mathrm{SO}(3,1)$. The configuration can be obtained by using U-duality. We start with the configuration described in section \ref{sec:dyonickk6} and given in table \ref{tab:braneconf3}. We interchange the directions 10 and 11 bringing the NUT direction in 11 and we perfom a double T-duality in direction 2 and 3. This leads to a configuration of massive IIA with a F1, a D6 and a D8 described in table \ref{tab:braneconfm1} where the direction 9 is the timelike one.

The $\slfour$ embedded in  E$_{11}$ corresponding to this solution is given by (see for instance \cite{Englert:2007qb}),
\begin{eqnarray}
h_1 &=& -\frac{1}{3}(K_{1}+...+K_{8})+\frac{2}{3}(K_{9}+K_{10}+K_{11}) \nonumber, \\
h_2 &=& \frac{1}{3}(K_{2}+K_{3}+K_{6}+K_{7}+K_{8}+K_{11})\\
&&\quad\quad\quad\quad\quad-\frac{2}{3}(K_{1}+K_{4}+K_{5}+K_{9}+K_{10}), \nonumber \\
h_3 &=& -\frac{1}{3}(K_{1}+K_{2}+K_{3}+K_{6}+K_{7}+K_{8})+\frac{2}{3}(K_{4}+K_{5}+K_{11}) \nonumber, 
\end{eqnarray}
for the Cartan elements and
\begin{eqnarray}
e_1 &=& R^{9 {10}11},\nonumber\\
\label{eqn:e11embeddingm1}
e_2 &=& R^{2 3 6 7 8 {11}}\\
e_3 & = &R^{4 5 {11}} \nonumber ,
\end{eqnarray}
for the positive simple step operators. The three branes in the configuration correspond to the non zero $\slfour$ step operators containing the time direction 9, namely:
\begin{eqnarray}
&\, & \mathrm{F1} :  e_1 = R^{9 {10}11} ,\qquad  (\rm{level \,  1}), \nonumber\\
&\, & \mathrm{D6} :  [e_1,e_2] = \frac{1}{3}R^{{11} | 2 3 6 7 8 9 {10}11}  \qquad  (\rm{level \, 3}),\nonumber\\
&\, & \mathrm{D8} : [[e_1,e_2],e_3] = -\frac18  R^{{11} | {11} | 2 3 4  5 6 7 8 9 {10}11}  \qquad (\rm{level\,  4}),
\end{eqnarray}
where we have used the conventions and normalizations of appendix E in \cite{Henneaux:2008nr}. We note also that the D$8$ brane is an exact solution of the $\sigma$-model in that the transverse space is one-dimensional and there is no issue of smearing or unsmearing the solution.

\begin{table}[h!]
\begin{center}
\begin{tabular}{|c|c|ccccccc|c|c|}
\hline
Branes &$1$& $2$ & $3$ & $4$ & $5$ & $6$ & $7$ & $8$ & $9$\, (t) & $10$\\
\hline\hline
F1 &\, & \, & \, &\,& \, & \, &\, &\, &$\bullet$ &$\bullet$ \\
\hline
D6 &\, &$\bullet$ &$\bullet$ &\,& \, & $\bullet$ &$\bullet$  &$\bullet$ &$\bullet$ &$\bullet$ \\
\hline
D8 &\, & $\bullet$ &$\bullet$ &$\bullet$& $\bullet$ &$\bullet$ &$\bullet$ &$\bullet$ &$\bullet$ &$\bullet$  \\
\hline
\end{tabular}
\end{center}
\caption{\small \sl \small The space-time positions of the three branes in the bound state between a 1F, a  D6 and a D8. Here 9 is time.}
\label{tab:braneconfm1}
\end{table}

\subsection{An $\slfour/\mathrm{SO}(2,2)$ $\sigma$-model}
\label{sec:sl4sigmamodelso22}

\begin{figure}
\begin{center}
\begin{overpic}[scale=0.6]{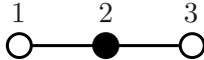}
\put(86,20){3}
\put(47,20){2}
\put(7,20){1}
\end{overpic}
\caption{\small \small Dynkin diagram for $A_3$, with the black node indicating time. In this case the fixed subalgebra under the temporal involution is $\mathfrak{so}(2,2)$, as by choosing time to be the middle node we get the involution (\ref{eqn:involutionso22}).}
\label{fig:so22}
\end{center}
\end{figure}

Let us now consider the other $\slfour$ case, with a $\sigma$-model defined by a $\mathrm{SO}(2,2)$ subgroup. First, consider the involution acting on the Lie algebra. The temporal involution defined by figure \ref{fig:so22} is given by
\begin{eqnarray}
\label{eqn:involutionso22}
\Omega(e_1) = -f_1, & \Omega(e_2) = f_2, & \Omega(e_3) = -f_3 ,
\end{eqnarray}
as in this case $e_2$ is the generator whose simple root is the node associated to time. In this case we define $k'_1,k'_2,k'_3,k'_{12},k'_{23}$ and $k'_{123}$ to be the six generators spanning the $\mathfrak{so}(2,2)$ subalgebra of $\mathfrak{sl}(4,\mathbb{R})$, and we let $s'_1,s'_2,s'_3,s'_{12}, s'_{23}$ and $s'_{123}$ together with $h_1,h_2$ and $h_3$ span its complement $\hat{\mathfrak{p}}$. We find hence the reductive decomposition 
\begin{equation}
\label{eqn:reductivesl4so22}
\mathfrak{sl}(4,\mathbb{R}) = \mathfrak{so}(2,2) \oplus \hat{\mathfrak{p}} .
\end{equation}
For more details on $\mathfrak{sl}(4,\mathbb{R})$ and its reductive decomposition with respect to $\mathfrak{so}(2,2)$ see appendix \ref{app:sl4rso22}. Up to the equations of motion this $\sigma$-model is identical to the previous $\slfour$-case. From the change of involution, the new equations of motion of the three Cartan fields become
\begin{eqnarray}
\partial_{\xi}^2 \phi_1 + \frac{1}{2}\left(-P_{\xi,1}^2+P_{\xi,4}^2+P_{\xi,6}^2\right) &=& 0,\nonumber \\
\partial_{\xi}^2 \phi_2 + \frac{1}{2}\left(P_{\xi,2}^2+P_{\xi,4}^2+P_{\xi,5}^2+P_{\xi,6}^2\right) &=& 0, \\
\partial_{\xi}^2 \phi_3 + \frac{1}{2}\left(-P_{\xi,3}^2+P_{\xi,5}^2+P_{\xi,6}^2\right) &=& 0\nonumber, 
\end{eqnarray}
and for the six coset field strengths
\begin{eqnarray}
\partial_{\xi} P_{\xi,1}+P_{\xi,2}P_{\xi,4}+P_{\xi,5}P_{\xi,6}+P_{\xi,1}(2\partial_{\xi}\phi_1-\partial_{\xi} \phi_2)&=&0, \nonumber\\
\partial_{\xi} P_{\xi,2}+P_{\xi,1}P_{\xi,4}-P_{\xi,3}P_{\xi,5}-P_{\xi,2}(\partial_{\xi}\phi_1-2\partial_{\xi}\phi_2+\partial_{\xi} \phi_3)&=&0,\nonumber\\
\label{eqn:slfoureqmsforPs2}
\partial_{\xi} P_{\xi,3}-P_{\xi,2}P_{\xi,5}-P_{\xi,4}P_{\xi,6}+P_{\xi,3}(2\partial_{\xi}\phi_3-\partial_{\xi} \phi_2) &=&0,\\
\partial_{\xi} P_{\xi,4}-P_{\xi,3}P_{\xi,6}+P_{\xi,4}(\partial_{\xi}\phi_1+\partial_{\xi} \phi_2-\partial_{\xi} \phi_3) &=&0,\nonumber\\
\partial_{\xi} P_{\xi,5}+P_{\xi,1}P_{\xi,6}-P_{\xi,5}(\partial_{\xi}\phi_1-\partial_{\xi} \phi_2-\partial_{\xi} \phi_3) &=&0,\nonumber\\
\partial_{\xi} P_{\xi,6}+P_{\xi,6}(\partial_{\xi}\phi_1+\partial_{\xi} \phi_3) &=&0.\nonumber
\end{eqnarray}
Apart from these equations we also have the lapse constraint
\begin{eqnarray}
0 &= &(\partial_{\xi} \phi_1)^2-\partial_{\xi} \phi_1\partial_{\xi} \phi_2+(\partial_{\xi} \phi_2)^2-\partial_{\xi} \phi_2\partial_{\xi} \phi_3+(\partial_{\xi} \phi_3)^2 \nonumber \\
&&+\frac{1}{4}(P_{\xi,1}^2-P_{\xi,2}^2+P_{\xi,3}^2-P_{\xi,4}^2-P_{\xi,5}^2-P_{\xi,6}^2) .
\end{eqnarray}

\subsubsection{A M2 $\subset$ M5$^2$ with magnetic Kaluza-Klein charge}
\begin{table}[!]
\begin{center}
\begin{tabular}{|c|ccc|cccc|c|c|c|c|}
\hline
Branes &$1$& $2$ & $3$ & $4$ & $5$ & $6$ & $7$ & $8$\, (t) & $9$ & $10$\, (N) & $11$\\
\hline\hline
M2 &\, & \, & \, &\,& \, & $\bullet$ &$\bullet$ &$\bullet$ &\, &\,& \, \\
\hline
M5 &\, & \, & \, &\,& \, & $\bullet$ &$\bullet$  &$\bullet$ &$\bullet$ &$\bullet$& $\bullet$ \\
\hline
M5 &\, & \, & \, & $\bullet$ & $\bullet$  & $\bullet$ &$\bullet$  &$\bullet$& \, &$\bullet$& \, \\
\hline
KK6 &\, & \, & \, &$\bullet$& $\bullet$ &$\bullet$ &$\bullet$ &$\bullet$ &$\bullet$ &$\bullet$& $\bullet$ \\
\hline
\end{tabular}
\end{center}
\caption{\small \sl \small The space-time positions of the four branes in the bound state between an M2, two M5 and a KK6. Observe that here 8 is time. Direction 10 is still the Taub-NUT direction.}
\label{tab:braneconf5}
\end{table}
\begin{figure}[h!]
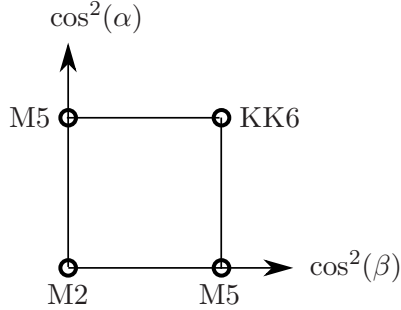

\begin{center}
\begin{overpic}[scale=0.8]{braneinterpolation.ps}
\put(90,4){$\cos^2(\beta)$}
\put(1,90){$\cos^2(\alpha)$}
\put(0,-6){M2}
\put(52,-6){M5}
\put(-13,55){M5}
\put(66,55){KK6}
\end{overpic}
\caption{\small The four different single brane limits of the M2$\subset$M5$^2 \subset$ KK6 bound state. Here the round circles indicate the pure brane limits, i.e. the four points $(0,0),(1,0),(0,1)$ and $(1,1)$. Both the $\cos^2(\beta) = 0$ and $\cos^2(\alpha) = 0$ axes correspond to a dyonic membrane.}
\label{fig:braneinterpolation2}
\end{center}
\end{figure}

Let us now repeat the procedure above for $\slfour/\sothreeone$ but in the $\mathrm{SO}(2,2)$ case. Hence we start with the Kaluza Klein monopole and act with a $\mathrm{SO}(2,2)$ transformation on it to generate some more general bound state. Again the KK6 is a highest weight, and the compact subgroup of $\mathrm{SO}(2,2)$, the product group $\mathrm{SO}(2) \times \mathrm{SO}(2)$, generate the two angle solution shown in figure \ref{fig:braneinterpolation2}. Note as now $e_2$ will be the generator with a time index, two generators at level 2 in $\mathfrak{e}_{11}$ will have a time index, namely both $e_{12}$ and $e_{23}$. This will give us two physical M5 branes. The corresponding harmonic functions are 
\begin{eqnarray}
H_1 &=& \sin^2(\beta) + \cos^2(\beta) H ,\\
H_2 &=& H, \\
H_3 &=& \sin^2(\alpha) + \cos^2(\alpha) H,
\end{eqnarray}
and one deduce the different limits by comparing to our different $\slthree$ $\sigma$-models.
This bound state is also described in \cite{Larsson:2001wt}.

\subsubsection{A (D4,D6$^2$,D8) bound state}

We now turn to the description of half-BPS states of massive Type IIA in the coset $\slfour/\mathrm{SO}(2,2)$. The configuration can be obtained again by using U-duality. We start with the configuration described above and displayed in table \ref{tab:braneconf5}. We interchange the directions 10 and 11 bringing the NUT direction to be 11 and we perfom a double T-duality in direction 2 and 3. This lead to a configuration of massive IIA with a D4, two D6 and a D8 described in table \ref{tab:braneconfm2}, where now the direction 8 is the timelike one.

The $\slfour$ embedded in  E$_{11}$ corresponding to this solution is the same as in section \ref{sec:f1d6d8} and  given  by (\ref{eqn:e11embeddingm1}). The only difference is that the involution is now given by (\ref{eqn:involutionso22}) selecting the time to be in direction $8$.

The four branes in the configuration correspond to the non zero $\slfour$ step operators containing the time direction 8, namely:
\begin{eqnarray}
&\, & \mathrm{D4} :  e_2 = R^{2 3 6 7 8 {11}} ,\qquad  (\rm{level \,  2}), \nonumber\\
&\, & \mathrm{D6} :  [e_1,e_2] = \frac{1}{3}R^{{11} | 2 3 6 7 8 9 {10}11}  \qquad  (\rm{level \, 3}),\nonumber\\
&\, & \mathrm{D6} :  [e_1,e_3] = \frac{1}{3}R^{{11} | 2 3 6 7 8 9 {10}11} ,\qquad  (\rm{level \,  3}), \nonumber\\
&\, & \mathrm{D8} : [[e_1,e_2],e_3] = -\frac18 R^{{11} | {11} | 2 3 4  5 6 7 8 9 {10}11}  \qquad (\rm{level\,  4}).
\end{eqnarray}

\begin{table}[h!]
\begin{center}
\begin{tabular}{|c|c|cccccc|c|cc|}
\hline
Branes &$1$& $2$ & $3$ & $4$ & $5$ & $6$ & $7$ & $8$\, (t) & $9$ & $10$\\
\hline\hline
D4 &\, &$\bullet$ &$\bullet$ &\,& \, & $\bullet$ &$\bullet$ &$\bullet$ &\, & \,  \\
\hline
D6 &\, & $\bullet$ &$\bullet$ &\,& \, & $\bullet$ &$\bullet$  &$\bullet$ &$\bullet$ &$\bullet$ \\
\hline
D6 &\, &$\bullet$ & $\bullet$ & $\bullet$ & $\bullet$  & $\bullet$ &$\bullet$  &$\bullet$& \, &\, \\
\hline
D8 &\, &$\bullet$& $\bullet$&$\bullet$& $\bullet$ &$\bullet$ &$\bullet$ &$\bullet$ &$\bullet$ &$\bullet$ \\
\hline
\end{tabular}
\end{center}
\caption{\small \sl \small The space-time positions of the four branes in the bound state between a D4, two D6 and a D8. Observe that here 8 is time.}
\label{tab:braneconfm2}
\end{table}

\subsection{Intersection rules}

Similar to what has been discussed in section~\ref{intersl3}, one can also consider marginal intersections of the more complicated bound states derived above using $\slfour$. In our algebraic description this requires finding commuting copies of $\slfour$ inside $\mathrm{E}_{11}$. The largest number of such commuting subgroups we have found is two, and the Dynkin diagram suggests that this is also the maximum possible. The intersection of bound states corresponding to $\slfour\times\slfour\subset\mathrm{E}_{11}$ preserves $1/4$ of the supersymmetry and, if realized only with branes, has two overall non-compact dimensions.

Both the case of a three brane bound state of section~\ref{sec:sl4sigmamodel} and the case of the four brane bound state of section~\ref{sec:sl4sigmamodelso22} can appear in the intersection. For simplicity, we only give the simple step operators generating the two $\slfour$ subgroups. The first one we take to be the same as in section~\ref{sec:sl4sigmamodel} (see (\ref{eqn:e11embedding2})):
\begin{eqnarray}
\label{eqn:3branesbound}
e_1 = R^{9\,10\,11}\,,\quad e_2= R^{6\,7\,8}\,,\quad e_3 = R^{4\,5\,10}\,.
\end{eqnarray}
This corresponds to a half-BPS bound state of three branes M$2\subset$M$5\subset$KK$6$, with NUT direction $10$. For the second bound state we take
\begin{eqnarray}
\label{eqn:sl4bound1}
e_1' = R^{4\,7\,9}\,,\quad e_2'= R^{5\,8\,11}\,,\quad e_3' = R^{3\,7\,10}\,,
\end{eqnarray}
corresponding to an $\text{M}2\subset\text{M}5\subset\text{KK}6$ bound state with NUT direction $7$. One can easily verify that the generators in (\ref{eqn:sl4bound1}) generate an $\slfour$ with subgroup $\mathrm{SO}(3,1)$. Time here was chosen in the direction $9$.

Alternatively, we could have taken for example the following generators
\begin{eqnarray}
\label{eqn:sl4bound2}
e_1' = R^{5\,8\,11}\,,\quad e_2'= R^{4\,7\,9}\,,\quad e_3' = R^{3\,8\,10}\,,
\end{eqnarray}
leading to an $\slfour$ with subgroup $\mathrm{SO}(2,2)$. This bound state consists of four branes and can equally well be intersected with the bound state corresponding to (\ref{eqn:3branesbound}). We take this as the final example of how one can easily construct BPS states and their intersections using algebraic considerations based on the extensions $\mathrm{E}_{10}$ and $\mathrm{E}_{11}$ of the hidden symmetry groups known in $D=11$ supergravity.

\vspace{10mm}
{\bf Acknowledgements}

\noindent
LH is a Senior Research Associate and AK is a Research Associate of the Fonds de la Recherche Scientifique-FNRS, Belgium. 
This work has been supported in part by IISN-Belgium (conventions 4.4511.06, 4.4505.86 
and 4.4514.08) and by the Belgian Federal Science Policy Office through the Interuniversity 
Attraction Pole P6/11. 

\appendix

\section{Supergravity actions and equations of motion}

Our conventions regarding forms are that a $p$-form $V$ is written in coordinates as $V = \frac{1}{p!}V_{{\tilde{\mu}_1}...{\tilde{\mu}_p}}\mathrm{d}x^{\tilde{\mu}_1}\wedge...\wedge \mathrm{d}x^{\tilde{\mu}_p}$ so that for example $\frac{1}{4!} \sqrt{-g} F^2 \mathrm{d}^{11}x= F \wedge \star F$. The Levi-Cevita tensor is defined such that $\epsilon_{\tilde{1} ... \tilde{10}\tilde{11}} = 1$.
\subsection{Eleven dimensional supergravity}
\label{app:supergravity}

The bosonic part of the Lagrangian for eleven dimensional supergravity is
\begin{equation}
\mathcal{L}_{11d} = \sqrt{-g}R-\frac{1}{2}F_4 \wedge \star F_4+\frac{1}{6}A_3\wedge F_4 \wedge F_4 .
\end{equation}
Here we have defined
\begin{equation}
\label{eqn:electricpotential}
F_4 = \mathrm{d}A_3 .
\end{equation}
The equation of motion for the 4-form field strength is
\begin{equation}
\mathrm{d}\big(\star F_4+\frac{1}{2} A_3 \wedge \mathrm{d}A_3\big) = 0 .
\end{equation}
If we define the dual field strength 
\begin{equation}
\label{eqn:duality}
F_7 = \star F_4,
\end{equation}
it is written in terms of a dual potential $A_6$ as
\begin{equation}
\label{eqn:sevenform}
F_7 = \mathrm{d}A_6-\frac{1}{2}A_3 \wedge \mathrm{d}A_3 .
\end{equation}
This way the equation of motion for $F_4$ become the Bianchi identity for $F_7$ as should be the case for a dual field strength. The Einstein equation is
\begin{equation}
R_{\tilde{\mu} \tilde{\nu}}-\frac{1}{12}F_{\tilde{\mu} \tilde{\rho_1} \tilde{\rho_2} \tilde{\rho_3}}{F_{\tilde{\nu}}}^{\tilde{\rho_1} \tilde{\rho_2} \tilde{\rho_3}}+\frac{1}{6 \cdot4!}g_{\tilde{\mu} \tilde{\nu}}F^2 = 0 .
\end{equation}

\section{Some details on $\mathfrak{sl}(n,\mathbb{R)}$ algebras}
\label{app:slnr}

In this appendix we perform the reductive decompositions of $\mathfrak{sl}(3, \mathbb{R})$ and $\mathfrak{sl}(4, \mathbb{R})$ necessary for the discussions in the main text. Recall that given a Lie algebra involution $\Omega : \mathfrak{g} \rightarrow \mathfrak{g}$ we can decompose the algebra as $\mathfrak{g} = \mathfrak{k} \oplus \mathfrak{p}$, where $\mathfrak{k}$ is the fixed point set under $\Omega$, and $\mathfrak{p}$ its complement, under the involution. This vector space decomposition is commonly called the reductive decomposition. As $[\mathfrak{k} ,\mathfrak{p} ]\subset \mathfrak{p}$ we can describe $\mathfrak{p}$ as a $\mathfrak{k}$ representation. We will use these representations in the main text and we will therefore describe them here explicitly for $\mathfrak{sl}(3, \mathbb{R})$ and $\mathfrak{sl}(4, \mathbb{R})$. One will then see that the highest weights of these representations can be associated to the extremal M2, M5 and KK6 branes and we use this fact when we determine their orbit spaces in $\slthree$ and $\slfour$.

\subsection{Reductive decomposition of $\mathfrak{sl}(3,\mathbb{R})$}
\label{app:sl3r}

The temporal involution $\Omega$ act as
\begin{eqnarray}
\Omega(h_i) = -h_i,& \Omega(e_1) = f_1, & \Omega(e_2) = -f_2 . 
\end{eqnarray}
Hence, defining
\begin{equation}
k_1 = \frac{e_1+f_1}{2},\quad k_2 = \frac{e_2-f_2}{2},\quad k_{12}= \frac{e_{12}+f_{12}}{2},
\end{equation}
and
\begin{equation}
s_1 = \frac{e_1-f_1}{2},\quad s_2 = \frac{e_2+f_2}{2},\quad s_3= \frac{e_{12}-f_{12}}{2},
\end{equation}
we have
\begin{equation}
\mathfrak{k} = \mathrm{Span}_{\mathbb{R}}\{ k_1,k_2,k_{12}\},
\end{equation}
together with its complement
\begin{equation}
\mathfrak{p} = \mathrm{Span}_{\mathbb{R}}\{ h_1,h_2, s_1,s_2,s_{12}\} .
\end{equation}
As $\mathfrak{k}$ is isomorphic to $\mathfrak{sl}(2,\mathbb{R})$ via 
\begin{eqnarray}
h =-4(k_1+k_2-k_{12}),&e= 2\sqrt{2}(k_2-k_{12}),&f =  -2\sqrt{2}(k_1+k_{2}),\nonumber \\
\end{eqnarray}
we can think of $\mathfrak{p}$ as a familiar $\mathfrak{sl}(2,\mathbb{R})$-representation, and it turns out to be the irreducible $\bf{5}$, with highest weight vector $\lambda =h_1+2s_1$. This $\lambda$ is exactly the M2 brane discussed in the main text. Note that $k_2$ is the compact generator of $\mathfrak{k}$.

\subsection{Decomposition of the $\mathfrak{sl}(4,\mathbb{R)}$ algebra - $\mathfrak{so}(3,1)$ case}
\label{app:sl4rso31}

\begin{figure}
\begin{center}
\begin{overpic}[scale=0.6]{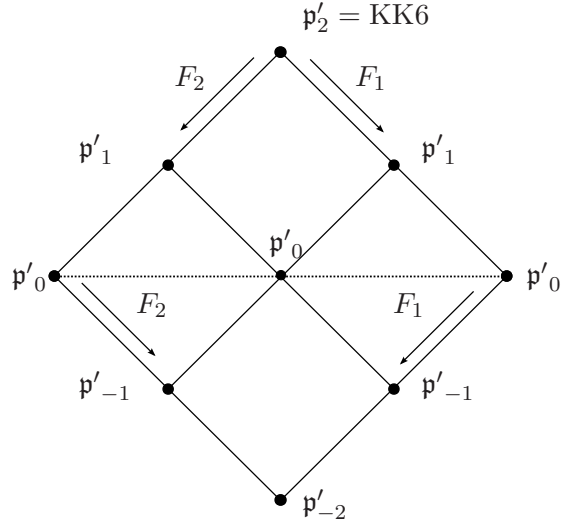}
\put(55,0){$\mathfrak{p}'_{-2}$}
\put(55,103){$\mathfrak{p}'_{2} = \mathrm{KK}6$}
\put(48,55){$\mathfrak{p'}_{0}$}
\put(102,48){$\mathfrak{p'}_{0}$}
\put(-6,48){$\mathfrak{p'}_{0}$}
\put(8,75){$\mathfrak{p'}_{1}$}
\put(80,75){$\mathfrak{p'}_{1}$}
\put(8,25){$\mathfrak{p'}_{-1}$}
\put(80,25){$\mathfrak{p'}_{-1}$}
\put(28,90){$F_2$}
\put(66,90){$F_1$}
\put(20,42){$F_2$}
\put(74,42){$F_1$}
\end{overpic}
\caption{\small Diagram over $\mathfrak{p'}$ as a  $\mathfrak{so}(3,1)$ representation. After the dotted line, the direction of $F_1$ and $F_2$ change. This implies for example that ${Ad_{F_2}}^4 = {Ad_{F_1}}^4$.}
\label{fig:repso31}
\end{center}
\end{figure}

Extending the above $\mathfrak{sl}(3, \mathbb{R})$ to $\mathfrak{sl}(4, \mathbb{R})$ we add a third $\mathfrak{sl}(2, \mathbb{R})$ algebra, $e_3, f_3$ and $h_3 = [e_3,f_3]$, commutating with the above given $\mathfrak{sl}(3, \mathbb{R})$ in such a way that $\mathfrak{sl}(4, \mathbb{R})$ is generated, i.e. with the extra positive step operators $e_{23}  =  [e_2,e_3] $ and $e_{123}  =  [e_1,[e_2,e_3]]$. The extra negative step operators are $f_{23}=-[f_2,f_3]$ and $f_{123} = -[f_1, [f_{23}]]$. We define the action of $\Omega$ as
\begin{eqnarray}
\Omega(e_1) = f_1, & \Omega(e_2)=-f_2, & \Omega(e_3) = -f_3. 
\end{eqnarray}
Hence we get $\mathfrak{so}(3,1)$ as fixed subalgebra, spanned by the six generators
\begin{align}
k_1 &= \frac{e_1+f_1}{2},&k_2 &= \frac{e_2-f_2}{2},& k_{12}&= \frac{e_{12}+f_{12}}{2},& \nonumber \\
k_3 &= \frac{e_3-f_3}{2},& k_{23} &= \frac{e_{23}-f_{23}}{2},&k_{123} &=  \frac{e_{123}+f_{123}}{2}.&
\end{align}
Defining
\begin{align}
s_1 &= \frac{e_1-f_1}{2},&s_2 &= \frac{e_2+f_2}{2},& s_{12}&= \frac{e_{12}-f_{12}}{2}, &\nonumber \\
s_3 &= \frac{e_3+f_3}{2},& s_{23} &= \frac{e_{23}+f_{23}}{2},&s_{123}& =  \frac{e_{123}-f_{123}}{2},&
\end{align}
we have
\begin{equation}
\mathfrak{p}' = \mathrm{Span}_{\mathbb{R}} (h_1,h_2,h_3,s_1,s_2,s_3,s_{12},s_{23},s_{123}) .
\end{equation}
Now, determining the $\mathfrak{so}(3,1)$ representation $\mathfrak{p}'$ constitutes we will diagonalize $\mathfrak{p}'$ with respect to a maximally commuting subalgebra of non-compact elements in $\mathfrak{so}(3,1)$. Since the reduced root system is of rank one, this subalgebra will in fact only contain one element, which we choose to be $H = 4k_{123}$. There are two pairs of creation and annihilation operators with respect to $H$, given by
\begin{eqnarray}
E_1 = 2(k_1+k_{23}), & F_1 = 2(-k_1+k_{23}) \nonumber, \\
E_2 = 2(k_3+k_{12}), & F_2 = -2(k_3-k_{13}) .
\end{eqnarray}
Observe that $[F_1, F_2] = 0$ and that $[E_1,E_2]=0$. With respect to $H$ we can decompose $\mathfrak{p}'$ in a sum of weight spaces
\begin{equation}
\label{eqn:so31weights}
\mathfrak{p}' = \mathfrak{p}'_{-2} \oplus \mathfrak{p}'_{-1} \oplus \mathfrak{p}'_{0} \oplus \mathfrak{p}'_{1} \oplus \mathfrak{p}'_{2},
\end{equation}
where
\begin{eqnarray}
\mathfrak{p}'_{-2} &=& \mathbb{R} \lambda_1 \nonumber, \\
\mathfrak{p}'_{-1} &= &\mathbb{R} \lambda_2 + \mathbb{R} \lambda_3 \nonumber ,\\
\mathfrak{p}'_{0} &=& \mathbb{R} \lambda_4 + \mathbb{R} \lambda_5 +\mathbb{R} \lambda_6, \\
\mathfrak{p}'_{1} &=& \mathbb{R} \lambda_7 + \mathbb{R} \lambda_8 \nonumber ,\\
\mathfrak{p}'_{2} &=& \mathbb{R} \lambda_9 \nonumber,
\end{eqnarray}
and the weights $\lambda_i$ are given as 
\begin{eqnarray}
\lambda_1 &=& -\frac{1}{2}(h_1+h_2+h_3)+s_{123} \nonumber, \\
\lambda_2 &=& s_{23}+s_1 \nonumber,  \\
\lambda_3 &=& s_{12}+s_3 \nonumber, \\
\lambda_4 &=& s_2 \nonumber,  \\
\lambda_5 &=& h_1+h_2-h_3 ,\\
\lambda_6 &=& h_1-h_2 -h_3 \nonumber,  \\
\lambda_7 &=&  s_{12}-s_3 \nonumber , \\
\lambda_8 &=& s_{23}-s_1  \nonumber , \\
\lambda_9 &=&  \frac{1}{2}(h_1+h_2+h_3)+s_{123} \nonumber.
\end{eqnarray}
The structure of this representation is shown in figure \ref{fig:repso31}. Observe that $\lambda_9$ is the extremal KK6 monopole, introduced in section \ref{sec:kk6}.

\subsection{Decomposition of the $\mathfrak{sl}(4,\mathbb{R)}$ algebra - $\mathfrak{so}(2,2)$ case}
\label{app:sl4rso22}

One can define the involution of $\mathfrak{sl}(4,\mathbb{R)}$ in another way, such that the fixed subalgebra instead becomes $\mathfrak{so}(2,2) \cong \mathfrak{sl}(2,\mathbb{R}) \times \mathfrak{sl}(2,\mathbb{R})$. Let $\Omega$ act on  $\mathfrak{sl}(4,\mathbb{R)}$ by (compare with (\ref{eqn:involutionso22}))
\begin{eqnarray}
\Omega(e_1) = -f_1, & \Omega(e_2)= f_2, & \Omega(e_3) = -f_3.
\end{eqnarray}
The difference from the above case is now that the `time-like' node is the middle one in the Dynkin diagram for $A_3$. The fixed subalgebra $\mathfrak{so}(2,2) $ is now spanned by the generators
\begin{align}
k'_1 &= \frac{e_1-f_1}{2},& k'_2 &= \frac{e_2+f_2}{2},& k'_{12}&= \frac{e_{12}+f_{12}}{2},& \nonumber \\
k'_3 &= \frac{e_3-f_3}{2},& k'_{23} &= \frac{e_{23}+f_{23}}{2},&k'_{123}& =  \frac{e_{123}+f_{123}}{2}.&
\end{align}
The complement $\tilde{\mathfrak{p}}$ is then spanned by
\begin{align}
s'_1 &= \frac{e_1+f_1}{2},\ & s'_2 &= \frac{e_2-f_2}{2},& s'_{12}&= \frac{e_{12}-f_{12}}{2}, &\nonumber \\
s'_3 &= \frac{e_3+f_3}{2},\ & s'_{23} &= \frac{e_{23}-f_{23}}{2},&s'_{123} &=  \frac{e_{123}-f_{123}}{2},&
\end{align}
together with $h_1, h_2$ and $h_3$ and we get the reductive decomposition
\begin{equation}
\mathfrak{sl}(4, \mathbb{R}) = \mathfrak{so}(2,2) \oplus \hat{\mathfrak{p}} .
\end{equation}
\begin{figure}
\begin{center}
\begin{overpic}[scale=0.6]{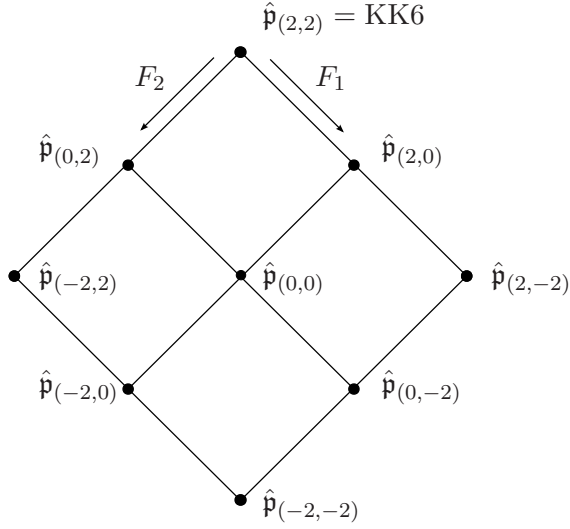}
\put(55,0){$\hat{\mathfrak{p}}_{(-2,-2)}$}
\put(55,103){$\hat{\mathfrak{p}}_{(2,2)} = \mathrm{KK}6$}
\put(55,48){$\hat{\mathfrak{p}}_{(0,0)}$}
\put(103,48){$\hat{\mathfrak{p}}_{(2,-2)}$}
\put(8,48){$\hat{\mathfrak{p}}_{(-2,2)}$}
\put(8,75){$\hat{\mathfrak{p}}_{(0,2)}$}
\put(80,75){$\hat{\mathfrak{p}}_{(2,0)}$}
\put(8,25){$\hat{\mathfrak{p}}_{(-2,0)}$}
\put(80,25){$\hat{\mathfrak{p}}_{(0,-2)}$}
\put(28,90){$F_2$}
\put(66,90){$F_1$}
\end{overpic}
\caption{\small Diagram over $\hat{\mathfrak{p}}$ as a  $\mathfrak{so}(2,2)$ representation. In contrast to the $\mathfrak{so}(3,1)$ case, both $F_1$ and $F_2$ square to zero.}
\label{fig:repso22}
\end{center}
\end{figure}
To determine the representation that  $\hat{\mathfrak{p}}$ constitues we use the isomorphism  $ \mathfrak{so}(2,2)  \cong \mathfrak{sl}(2,\mathbb{R}) \times \mathfrak{sl}(2,\mathbb{R}) $. We define the two commuting $ \mathfrak{sl}(2,\mathbb{R})$ by
\begin{eqnarray}
H_1 = 2(k'_2+k'_{123}), & E_1 = k'_1+k'_3-k'_{12}+k'_{23}, & F_1 = -k'_1-k'_3-k'_{12}+k'_{23} \nonumber \\
\end{eqnarray}
and
\begin{eqnarray}
H_2 = 2(-k'_2+k'_{123}), & E_2 = -k'_1+k'_3-k'_{12}-k'_{23}, & F_2 = k'_1-k'_3-k'_{12}-k'_{23} .\nonumber \\
\end{eqnarray}
Under these two sub-algebras we find the nine-dimensional representation
\begin{equation}
\hat{\mathfrak{p}} = ({\bf 3,3})
\end{equation}
where ${\bf 3}$ is the usual $\mathfrak{sl}(2,\mathbb{R})$ representation $(-2,0,2)$. By doing a decomposition with respect to the eigenspaces of $H_1$ and $H_2$, as in (\ref{eqn:so31weights}), we have
\begin{equation}
\label{eqn:so22weights}
\hat{\mathfrak{p}} = \sum_{n,m} \hat{\mathfrak{p}}_{(n,m)}
\end{equation}
where $n,m$ take the values $2,0$ and $2$. Again the highest weight of this representation, spanning $\hat{\mathfrak{p}}_{(2,2)}$ is the vector
\begin{equation}
\lambda = \frac{1}{2}(h_1+h_2+h_3)+s'_{123}
\end{equation}
corresponding to a KK6 monopole in the terminology of the main text. The $({\bf 3,3})$ representation is illustrated in figure \ref{fig:repso22} using the decomposition (\ref{eqn:so22weights}).

 \addcontentsline{toc}{section}{References}
\bibliographystyle{utphys}
\bibliography{Refdata}

\end{document}